\documentclass{aa}
\usepackage{txfonts}
\usepackage{graphicx}
\usepackage{array}
\usepackage{caption}
\usepackage{subcaption}
\usepackage{multirow}
\usepackage{amsmath}
\usepackage{hyperref}
\usepackage{enumitem}
\usepackage{soul}
\hypersetup{
    colorlinks=true,
    linkcolor=blue,
    urlcolor=cyan,      
    citecolor=magenta,
    }
\usepackage{natbib}
\bibpunct{(}{)}{;}{a}{}{,}

\begin{document}

\title{Diversities and similarities exhibited by multi-planetary systems and their architectures}
\subtitle{II. Radii of singles and multis}
\titlerunning{II. Radii of singles and multis}
\author{Alexandra Muresan\inst{1}
\and Carina M. Persson\inst{2}
\and Malcolm Fridlund\inst{2,3}
}
\institute{Dept. of Space, Earth and Environment, Chalmers University of Technology, Chalmersplatsen 4, 412 58 Gothenburg, Sweden.
\and Dept. of Space, Earth and Environment, Chalmers University of Technology, Onsala Space Observatory, 439 92 Onsala, Sweden.
\and Leiden Observatory, University of Leiden, PO Box 9513, 2300 RA, Leiden, The Netherlands.
}
\date{}

\abstract
{The discovered planets in apparent single-planet systems (hereafter singles) and those in systems with multiple detected planets (henceforth multis) exhibit a rich diversity of physical and orbital properties. 
However, it is still debated whether the observed singles and multis originate from the same underlying planet population. 
We investigate the differences and similarities between 1730~singles and 1522 multis in a catalogue of confirmed transiting planets orbiting main-sequence stars with spectral classes ranging from late-M to late-F. 
After we removed the hot Jupiters, the planet types and their fractional numbers were similar for the multis and singles hosted by FGK-type stars. 
Furthermore, the median radii of both the singles and the multis increase with host star temperature already from late- to early-type M dwarfs and further up to F~stars. 
Our analyses show that the singles are larger on average than the multis in our F, G, K, and M~samples. However, after we excluded the hot Jupiters, the radius distributions of the singles and multis orbiting FGK~stars are statistically indistinguishable, particularly at \mbox{$R < 4\, R_\oplus$}. 
In the FGK~sample, we also identified an unexpected and significant overabundance of multis, compared to singles, at radii of \mbox{$\approx1.4-1.6\, R_\oplus$}. 
For the early- and late-type M~samples, our work indicates that the multis are smaller on average than the singles and that the radius distributions of the multis and singles are different, except for the planets with \mbox{$R < 4\, R_\oplus$} hosted by early-type M~dwarfs. Nevertheless, these results for the two M~samples are inconclusive because the sample sizes of 167 and 101~planets are limited. 
In conclusion, our analyses reveal that the singles and multis, excluding the hot Jupiters, orbiting FGK~stars are overall consistent with originating from the same underlying population based on their planet types and radii. 
This interpretation is, however, not applicable within the region of the overabundance of multis at \mbox{$\approx 1.4-1.6\, R_\oplus$} identified in this work, which is an intriguing feature warranting further investigations.}

\maketitle

\section{Introduction} \label{sect:introduction}
Over the past few years, exoplanetary science has undergone significant progress, driven by technological advances and the detection of more than 6000~planets beyond the Solar System. 
One of the primary discoveries has been the rich diversity of exoplanets and planetary system architectures. Considerable effort has been devoted to characterising the observed exoplanet population and to developing theoretical models for their formation and evolution \citep[e.g.][]{mulders24, armitage, burn}. 

In the first paper of this series, we found that planets in multi-planetary systems (hereafter multis) often exhibit intra-system similarities in their radii, masses, and orbital spacings \citep{muresan}. These features stand in contrast to those of the systems with only one detected planet (hereafter singles). 
However, observational biases and selection effects limit our knowledge of the true multiplicity of the planets in a system and of their physical and orbital properties. It remains uncertain whether the observed singles are intrinsically single-planet systems or members of higher-multiplicity systems with undetected planets \citep[e.g.][]{weiss18b, zhu}.

Previous analyses of \textit{Kepler} systems identified a feature now denoted as the \textit{Kepler} dichotomy: an excess of single-planet systems compared to predictions based on the observed multiplicity distribution of multis \citep{lissauer11, ballard}. Several hypotheses have been proposed in order to explain the origin of this dichotomy. One possibility is that a substantial fraction of the single-planet systems are intrinsically singles and represent a distinct planet population.  
Another explanation is that the singles belong to a population of dynamically hot multi-planetary systems with high eccentricities and elevated mutual inclinations. This feature reduces the probability of detecting multiple planets in the same systems and increases the observed singles-to-multis ratio \citep[e.g.][]{ballard, zhu, he}. 
Corroborating this hypothesis, \citet{gilbert} identified that the eccentricity-radius relations of multis and singles are similar, thereby suggesting that they have a common origin. 
This idea is also supported by previous studies that found the properties of the \textit{Kepler} host stars of singles and multis to be statistically indistinguishable \citep{weiss18b, zhu}. 
Alternatively, the \textit{Kepler} dichotomy might arise due to a reduced transit detection efficiency for additional planets in the same system \citep{zink}. 

If singles and multis originate from the same underlying population, their radius distributions are expected to be similar. 
Analysing a sample of \textit{Kepler} planet candidates orbiting FGK-type stars, \citet{zhu} identified similar radius distributions for singles and multis, particularly for planets with radii $(R)<2\, R_\oplus$. Equivalently, \citet{weiss18b} reported statistically indistinguishable radius distributions of the singles and multis with orbital periods \mbox{$(P)>3$ days} and $R<\!4\, R_\oplus$ in a sample of \textit{Kepler} candidates hosted by FGK~stars. In contrast, \citet[][hereafter \color{blue}Lib24]{liberles} found significant differences between the radius distributions of the multis and singles orbiting M and late-type K dwarfs. 

Despite these prior studies, it remains unclear whether the radius distributions of singles and multis are similar across the full range of planet types and host star classes. 
The main objective of our work is to examine the detected transiting singles and multis and to test whether they are consistent with originating from the same underlying population based on their radii. 
In \citet{muresan}, we reported that the planetary radius distributions in systems with three or more planets are statistically indistinguishable. 
In this paper, we extend our previous analyses and also include systems with one and two detected planets. Particularly, we compare the radius distributions, planet types, and relative occurrences of the transiting singles and multis for different host star types.   
The data samples are described in Sect.~\ref{sect:samples}, followed by the statistical analyses in Sect.~\ref{sect:comparison}. We discuss our results in Sect.~\ref{sect:discussion}, before presenting the conclusions in Sect.~\ref{sect:conclusion}. 

\begin{table*}
\caption{Properties of singles and multis orbiting different types of stars.}
\label{tab:all_stypes}
\centering
\begin{tabular}
{cccccccc}
\hline\hline
\noalign{\smallskip}
Spec. & $T_\mathrm{eff}$ & $N_\mathrm{s}$ & $N_\mathrm{m}$ & $N_\mathrm{s}$ & $N_\mathrm{m}$ & $\widetilde{R_s}$ [$R_\oplus$] & $\widetilde{R_m}$ [$R_\oplus$]\\
type & [K] & & & excl. HJs & excl. HJs & excl. HJs & excl. HJs\\
\noalign{\smallskip}
\hline
\noalign{\smallskip}
F & 5960-6200 & 239 & 227 & 163 & 226 & 2.25±0.04 & 2.26±0.03\\
G & 5325-5960 & 814 & 644 & 621 & 634 & 2.36±0.02 & 2.29±0.02\\
K & 3890-5325 & 534 & 526 & 448 & 519 & 2.09±0.02 & 2.06±0.02\\
M0-2 & 3450-3890 & 87 & 80 & 76 & 80 & 1.94±0.04 & 1.64±0.04\\
M3-9 & 2310-3450 & 56 & 45 & 48 & 45 & 1.53±0.05 & 1.22±0.03\\
\hline
\end{tabular}
\tablefoot{$N$ and $\widetilde{R}$ represent the number and the median radii, respectively, of the singles (s) and multis (m). 
The host stars are divided into five spectral types based on their effective temperatures $T_\mathrm{eff}$.
The tabulated uncertainties are computed via Monte Carlo simulations, assuming a Gaussian distribution for the radius of each planet (Sect.~\ref{sect:radii}). 
Spectral types M0-2 and M3-9 correspond to early- and late-type M stars, respectively. The total of planets orbiting F-, G-, and K-type stars comprise the FGK~sample analysed in this work. "Excluding hot Jupiters" (excl. HJs) refer to the samples after removing the hot Jupiters ($P < 10$~days and $R > 6\, R_\oplus$).}
\end{table*}

\begin{figure}[htbp]
    \centering
    \begin{subfigure}{0.49\textwidth}
        \centering
        \includegraphics[width=\textwidth]{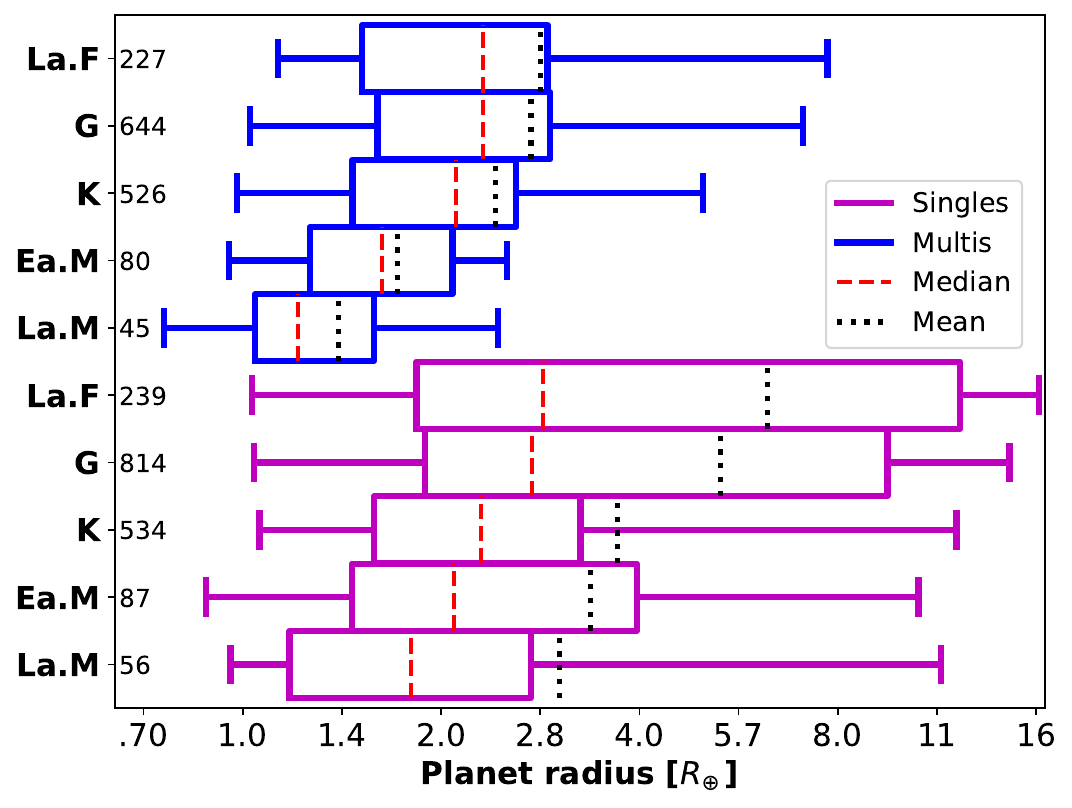}
        \caption{Full samples.}
        \label{fig:boxplots_a}
    \end{subfigure}
    %\par\medskip
    \begin{subfigure}{0.49\textwidth}
        \centering
        \includegraphics[width=\textwidth]{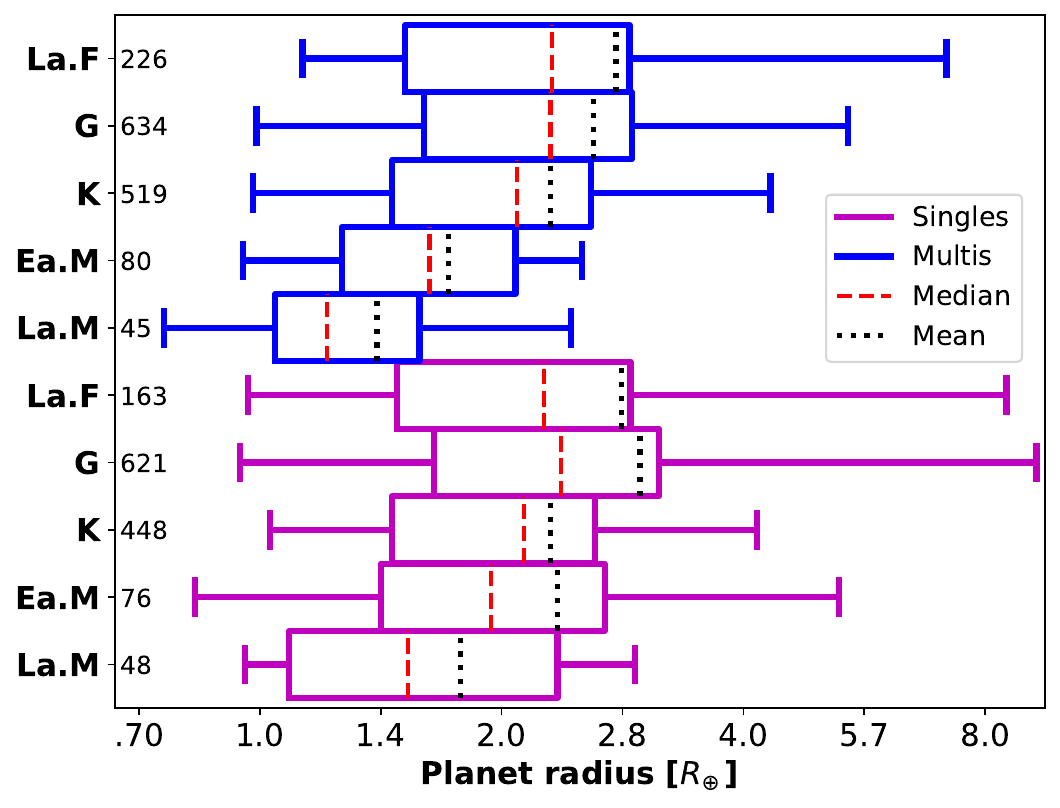}
        \caption{Full samples, excluding hot Jupiters.}
        \label{fig:boxplots_b}
    \end{subfigure}
    %\par\medskip
    \begin{subfigure}{0.49\textwidth}
        \centering
        \includegraphics[width=\textwidth]{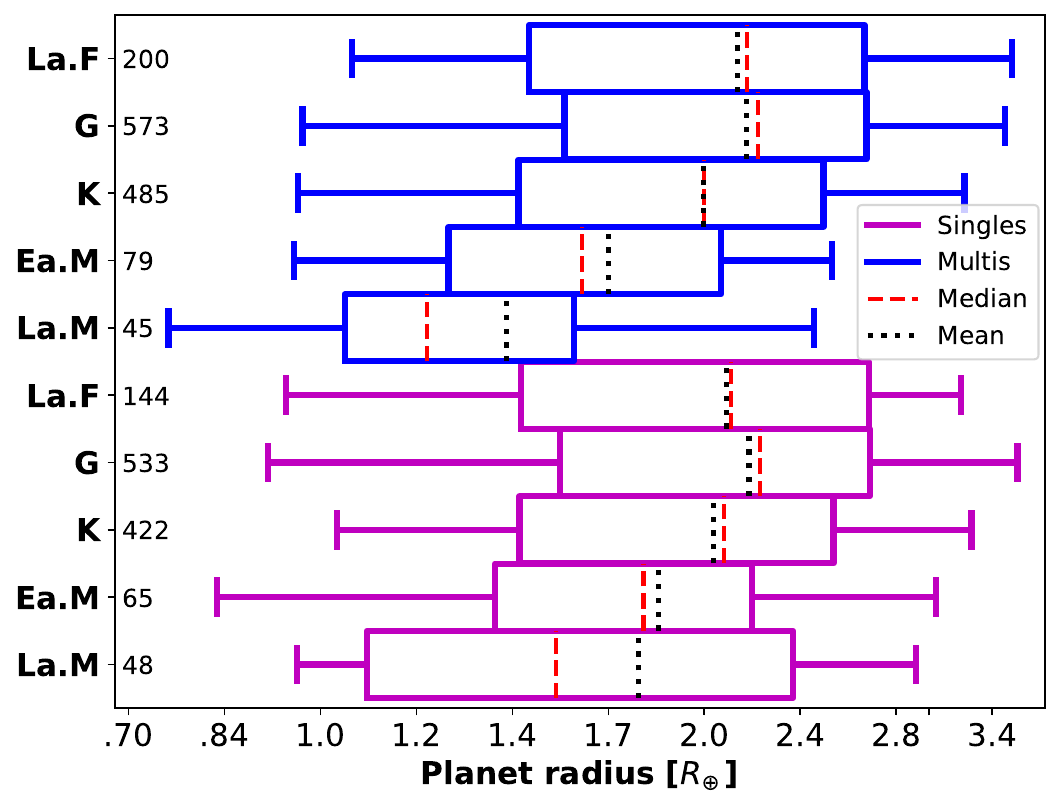}
        \caption{Planets with $R < 4\, R_\oplus$.}
        \label{fig:boxplots_c}
    \end{subfigure}
    \caption{Boxplots showing the radius distributions of all the singles (magenta) and multis (blue) examined in this work, divided into five samples based on the spectral type of the host star. Each box includes the data between the 0.25th and 0.75th quantiles, while the whiskers extend to the 5th and 95th percentiles. The median and mean radii of both the multis and singles increase as the host star temperature increases. The number of planets in each sample is indicated next to the host star type.}
    \label{fig:boxplots}
\end{figure}

\section{Data samples} \label{sect:samples}

We constructed our initial catalogue using the NASA Exoplanet Archive \citep{christiansen} by selecting all the confirmed planets with measured orbital periods and radii that orbit host stars in single-star systems with effective temperatures between 2310 and 6200~K.\footnote{Accessed at \url{https://exoplanetarchive.ipac.caltech.edu}. Our planet catalogue was last updated on 2025-08-15.} These limits correspond to stars spanning spectral types from late-M to late-F.

The NASA Exoplanet Archive classifies a planet as confirmed when it has been statistically validated in a refereed publication and found unlikely to be a false positive. As a result, our sample includes (i) statistically validated planets detected exclusively via the transit method, and (ii) planets confirmed through transits and complementary detection techniques. We adopted the Archive classification scheme and regard all planets in our catalogue as confirmed. 

We further restricted our sample to planets with \mbox{$R < 24\, R_\oplus$} that have reported uncertainties in both radius and orbital period, hosted by main-sequence stars with ages greater than 0.1~Gyr. This selection yielded a catalogue containing a large number of both singles and multis, comprising exclusively confirmed planets, thereby enhancing the overall reliability of the dataset. For planets with multiple radius measurements listed in the Archive, we adopted either the most recent value or the measurement with the smallest reported uncertainty.

Since our catalogue combines data from multiple missions and relies on heterogeneous methods for deriving the stellar properties, it is subject to intrinsic data inhomogeneities. This feature introduces certain caveats, which are further addressed in Sect.~\ref{sect:subsets}, where we analyse three additional, and more homogeneous, planet subsets. The results derived from our analyses of the individual subsets are mutually consistent and also agree with our analyses of the full catalogue. 

We divided our catalogue into five sub-samples based on the host star effective temperature and the corresponding spectral type, following the classification of \citet{mamajek}. Within each sub-sample, planets were further categorised as singles or multis according to the observed planet multiplicity in their host system. Table~\ref{tab:all_stypes} lists the adopted temperature ranges of the host stars as well as the numbers and median radii of the singles and multis in each sub-sample.

\begin{figure*}
\centering
\sidecaption
{\includegraphics[width=12cm]{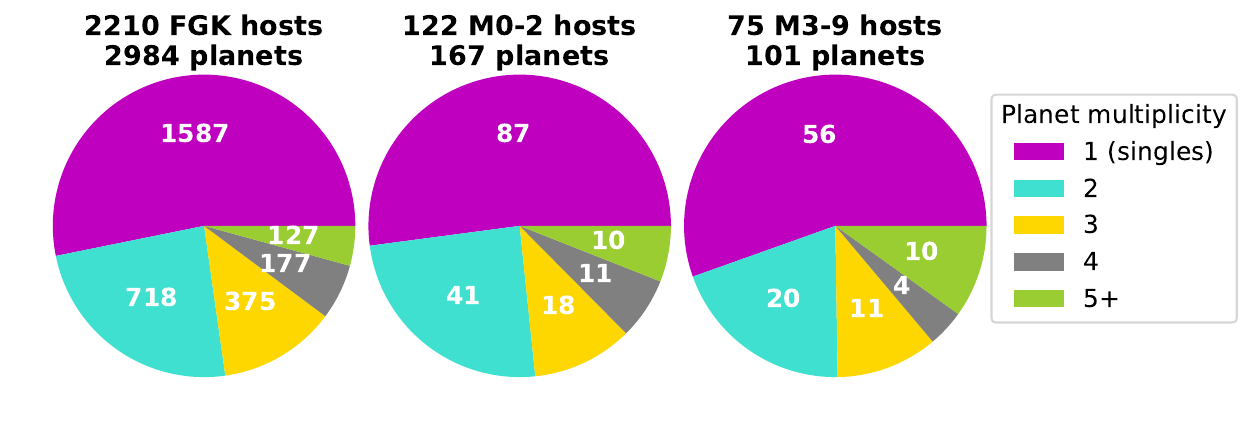}}
\caption{Three main planet samples, divided into singles (magenta) and multis, for which the colours indicate the observed planet multiplicity in the system. The corresponding number of planets is written in white font. 
M0-2 and M3-9 hosts represent early- and late-type M stars, respectively. 
The FGK~sample contains all the planets in Table~\ref{tab:all_stypes} that orbit F, G, and K-type stars.}
\label{fig:samples}
\end{figure*}

In order to enable robust comparisons between the confirmed multis and singles, we excluded from the initial catalogue only a few planets due to their very large radius uncertainties ($\sigma_R$). Imposing an overly restrictive upper bound on the relative radius errors ($\sigma_R/R$) might bias the resulting samples by preferentially removing unequal fractions of singles and multis and by favouring certain types of planets. 
We therefore applied a high upper limit of $\sigma_R/R < 15\%$, while the remaining radius uncertainties were accounted for in our analyses via Monte Carlo simulations, as explained in Sect.~\ref{sect:radii}. This selection retained more than 95\% of the planets orbiting M- and K-type dwarfs, as well as more than 90\% of the planets hosted by G and late-type F stars. 

The final sample contains 3252~transiting planets, comprising 1730~singles and 1522~multis in a total of 677~multi-planetary systems. Among the singles, 514 ($\approx 30\%$) were detected via radial velocity (RV) observations as well, while 484 ($\approx 32\%$) of the multis also have RV or transit-timing variation (TTV) measurements, with 264 and 275 planets possessing RV and TTV data, respectively. This implies that approximately 31\% of the planets in our catalogue are dynamically confirmed, while the remaining $\approx\!69\%$ were statistically validated and listed as confirmed in the NASA Exoplanet Archive. For simplicity, we refer to all the planets in our catalogue as confirmed. 

Figure \ref{fig:boxplots} shows the box plots of the planetary radius distributions in each of the five sub-samples associated with the host star spectral type. These plots reveal apparent differences in both the central values and the dispersions of the radii between singles and multis across various host star types.  
Notably, hot Jupiters, which are defined here as planets with \mbox{$P<10$ days} and \mbox{$R > 6\, R_\oplus$}, are predominantly singles orbiting host stars with \mbox{$T_\mathrm{eff}>5000$ K}. 
After we excluded this type of planet from our catalogue, the radius distributions of the singles and multis became very similar within each sub-sample, as seen in Fig.~\ref{fig:boxplots_b}. 
The near resemblance between the radii of the singles and multis in the F, G, and K-type sub-samples was further enhanced when we restricted the analysis to the small planets with \mbox{$R<4\, R_\oplus$}, as displayed in Fig.~\ref{fig:boxplots_c}. We stress, however, that similarities in the planet radii do not necessarily imply similar planetary masses or bulk compositions \citep[e.g.][]{leleu, schulze}.

As displayed in Fig.\ref{fig:boxplots}, following the removal of hot Jupiters, the F, G, and K sub-samples exhibit similar radii box plots. We therefore combined these sub-samples into one, henceforth denoted as the FGK~sample. In contrast, the planets in the two M~samples possess smaller radii on average and were therefore not included in the FGK~sample. Moreover, we analysed each M sample separately since the planets orbiting late-type M dwarfs (M3-9), particularly the multis, possess smaller median radii than the planets hosted by early-type M stars (M0-2), as shown in Fig.~\ref{fig:boxplots}. 

The number of systems and the observed planet multiplicities in the FGK, early-M, and late-M samples are displayed in Fig.~\ref{fig:samples}. All three samples exhibit similar multiplicity distributions with the singles constituting slightly more than 50\% of the population, followed by the two-planet systems and comparable fractions of the  three-, four-, and at least five-planet systems.

As shown in Table~\ref{tab:all_stypes} and Fig.~\ref{fig:boxplots}, hotter stars generally host larger planets, as previously reported by \citet{lozovsky}. Our work further shows that the median and mean radii of both the singles and the multis increase with host star temperature, even after the hot Jupiters were excluded. 
In agreement with prior research, our analyses also indicate that planets orbiting M dwarfs generally have smaller radii than planets hosted by FGK stars \citep[e.g.][]{gaidos}. Additionally, we found that they also tend to possess shorter orbital periods compared to planets around FGK stars, and that late-type M dwarfs host smaller planets on average than early-type M dwarfs (Sect.~\ref{sect:radii}).

\section{Comparisons between singles and multis} \label{sect:comparison}
\subsection{Percentages of different planet types} \label{sect:pct_pltypes}

\begin{figure}
\resizebox{\hsize}{!}{\includegraphics{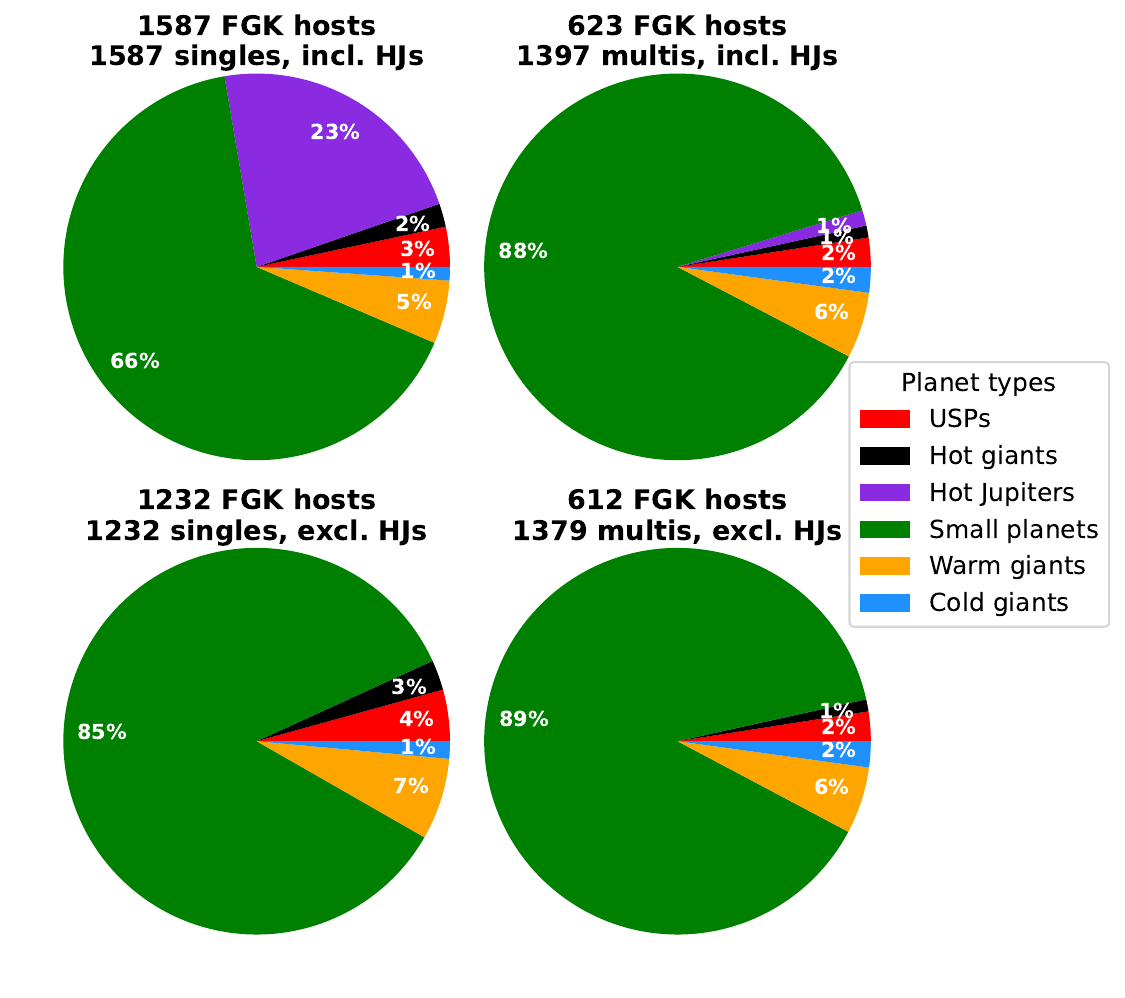}}
\caption{Singles and multis orbiting the FGK host stars in our catalogue, divided into six different planet types as indicated in the figure legend and Table~\ref{tab:pct_pltypes}. 
Upper panel: Entire FGK sample, including hot Jupiters, of which only 1\% are multis. 
Lower panel: FGK sample after removing the hot Jupiters. The singles and multis now have similar fractional numbers of the remaining five planet types.}
\label{fig:pltypes_fgk}
\end{figure}

\begin{figure}
\resizebox{\hsize}{!}{\includegraphics{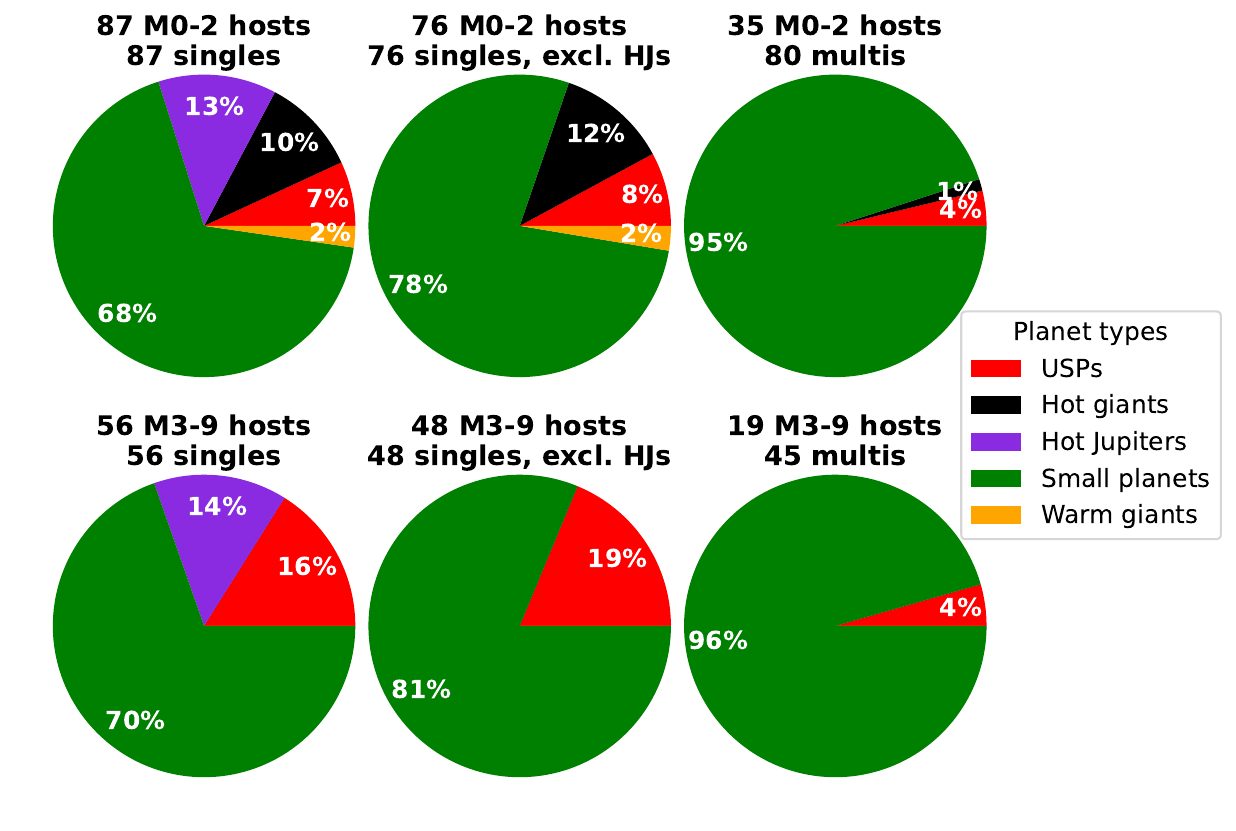}}
\caption{Singles and multis in the early-type (upper panel) and late-type (lower panel) M~sample, divided into five different planet types as listed in the figure legend and Table~\ref{tab:pct_pltypes}. The middle column displays the distribution of the singles after removing the hot Jupiters.}
\label{fig:pltypes_m}
\end{figure}

In order to assess whether single- and multi-planet systems host the same types of planets, we classified the singles and multis independently into six categories. The resulting fractional numbers, along with their associated Poisson uncertainties, are reported in Table~\ref{tab:pct_pltypes}: ultra-short-period planets (USPs; \mbox{$P \leq 1$ day,} \mbox{$R < 4\, R_\oplus$}), hot giants (\mbox{$P < 10$ days,} \mbox{$4 \leq R \leq  6\, R_\oplus$}), hot Jupiters (\mbox{$P < 1$ day,} \mbox{$R > 6\, R_\oplus$}), small planets (\mbox{$P > 1$ day,} \mbox{$R < 4\, R_\oplus$}), warm giants (\mbox{$10 \leq P \leq 100$ days,} \mbox{$R \geq 4\, R_\oplus$}), and cold giants (\mbox{$P > 100$ days,} \mbox{$R \geq 4\, R_\oplus$}). 
The distribution of planet types in the FGK and in the early- and late-type M samples are also illustrated in Figs.~\ref{fig:pltypes_fgk} and \ref{fig:pltypes_m}. 
In the FGK~sample, hot Jupiters account for 23\% of all the singles but only 1\% of the multis. When this category of planets is excluded, the fractions of the remaining five planet types comprising the singles are similar to those for the multis. 

\begin{table*}
\centering
\caption{Percentages of the six planet types, comprising the singles and multis.}
\label{tab:pct_pltypes}
\begin{tabular}
{cccccccc}
\hline\hline \noalign{\smallskip}
Spec. & s/m & USPs & Hot giants & Hot Jupiters & Small planets & Warm giants & Cold giants\\
type & & (\mbox{$P \leq 1$,} & (\mbox{$P < 10$,} & (\mbox{$P < 10$,} & (\mbox{$P > 1$,} & (\mbox{$10\leq P \leq 100$,} & (\mbox{$P > 100$,}\\
& & $R < 4$) & \mbox{$4\leq R \leq 6$}) & \mbox{$R > 6$}) & \mbox{$R < 4$}) & \mbox{$R \geq 4$}) & \mbox{$R \geq 4$})\\
\noalign{\smallskip}
\hline
\noalign{\smallskip}
\multirow{2}{*}{FGK} & s & 3±0.5\% & 2±0.4\% & 23±1\% & 66±2\% & 5±0.6\% & 1±0.3\%\\
& m & 2±0.4\% & 1±0.3\% & 1±0.3\% & 88±3\% & 6±0.6\% &  2±0.4\% \\
\hline
\noalign{\smallskip}
\multirow{2}{*}{M0-2} & s & 7±3\% & 10±3\% & 13±4\% & 68±9\% & 2±2\% & 0\% \\
& m & 4±2\% & 1±1\% & 0\% & 95±11\% & 0\% & 0\%\\
\hline
\noalign{\smallskip}
\multirow{2}{*}{M3-9} & s & 16±5\% & 0\% & 14±5\% & 70±11\% & 0\% & 0\%\\
& m & 4±3\% & 0\% & 0\% & 96±15\% & 0\% & 0\% \\
\hline
\end{tabular}
\tablefoot{Singles (s) and multis (m) in each of the FGK-type, early-type M (M0-2) and late-type M (M3-9) samples, divided into six different planet types, along with the corresponding percentages and Poisson uncertainties.
The radii ($R$) and orbital periods ($P$) are in units of Earth radii and days, respectively.
For instance, small planets with \mbox{$P > 1$ day} and \mbox{$R < 4\, R_\oplus$} account for $\approx66\%$ of the singles and $\approx88\%$ of the multis in the FGK~sample.}
\end{table*}

\begin{table*}
\centering
\caption{Percentages of the singles in different sub-samples.}
\label{tab:pct_singles}
\begin{tabular}
{cccccccc}
\hline\hline
\noalign{\smallskip}
Spec. type & Total & USPs & Hot giants & Hot Jupiters & Small planets & Warm giants & Cold giants\\
\noalign{\smallskip}
\hline
\noalign{\smallskip}
FGK & 53±1\% & 61±5\% & 69±7\% & 95±3\% & 46±1\% & 52±4\% & 38±7\%\\
M0-2 & 52±4\% & 67±17\% & 90±16\% & 100\% & 44±4\% & 100\% & 0\% \\
M3-9 & 55±5\% & 82±15\% & 0\% & 100\% & 48±6\% & 0\% & 0\% \\
\hline
\end{tabular}
\tablefoot{Percentages and binomial uncertainties of the detected planets that are singles in each of the sub-samples listed in Table~\ref{tab:pct_pltypes}. For instance, singles account for $\approx46\%$ of the confirmed small planets with \mbox{$P > 1$ day} and \mbox{$R < 4\, R_\oplus$} in the FGK~sample. 
Spectral types M0-2 and M3-9 represent early- and late-type M host stars, respectively.} 
\end{table*}

\begin{figure}[htbp]
    \centering
    \begin{subfigure}{0.49\textwidth}
        \centering
        \includegraphics[width=\textwidth]{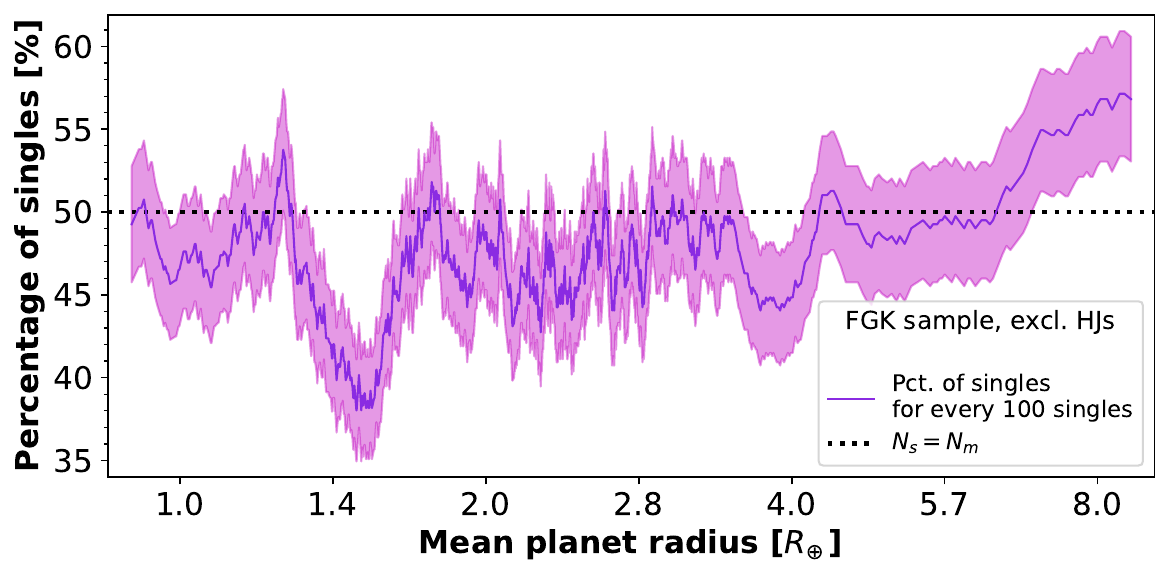}
        \caption{The FGK~sample, excluding the hot Jupiters.}
        \label{fig:pct_singles_a}
    \end{subfigure}
    \par\medskip
    \begin{subfigure}{0.49\textwidth}
        \centering
        \includegraphics[width=\textwidth]{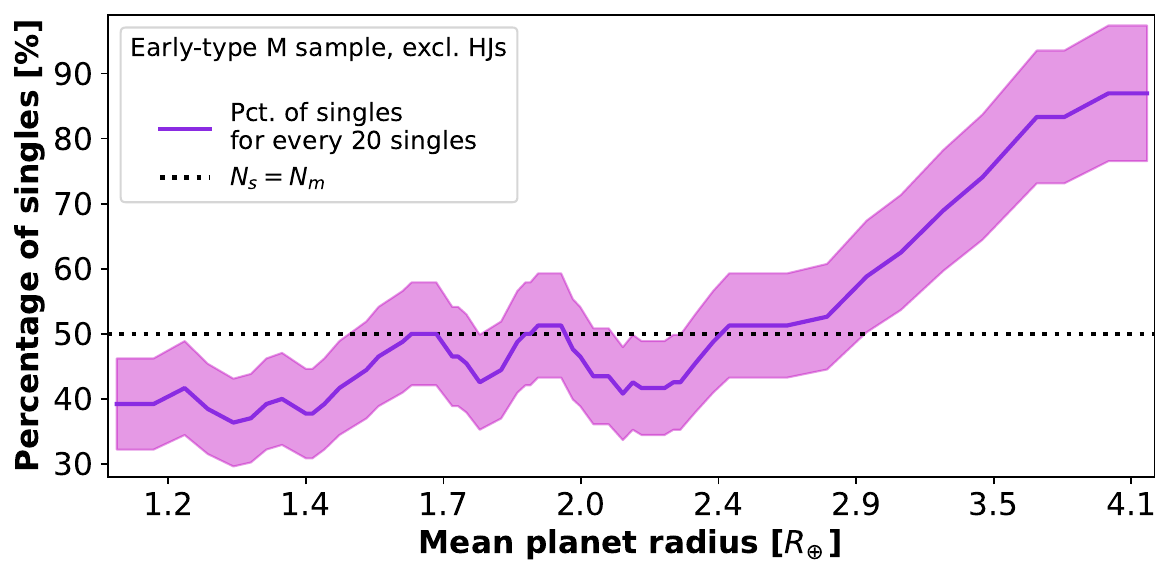}
        \caption{The early-type M sample, after removing the hot Jupiters.}
        \label{fig:pct_singles_b}
    \end{subfigure}
    \par\medskip
    \begin{subfigure}{0.49\textwidth}
        \centering
        \includegraphics[width=\textwidth]{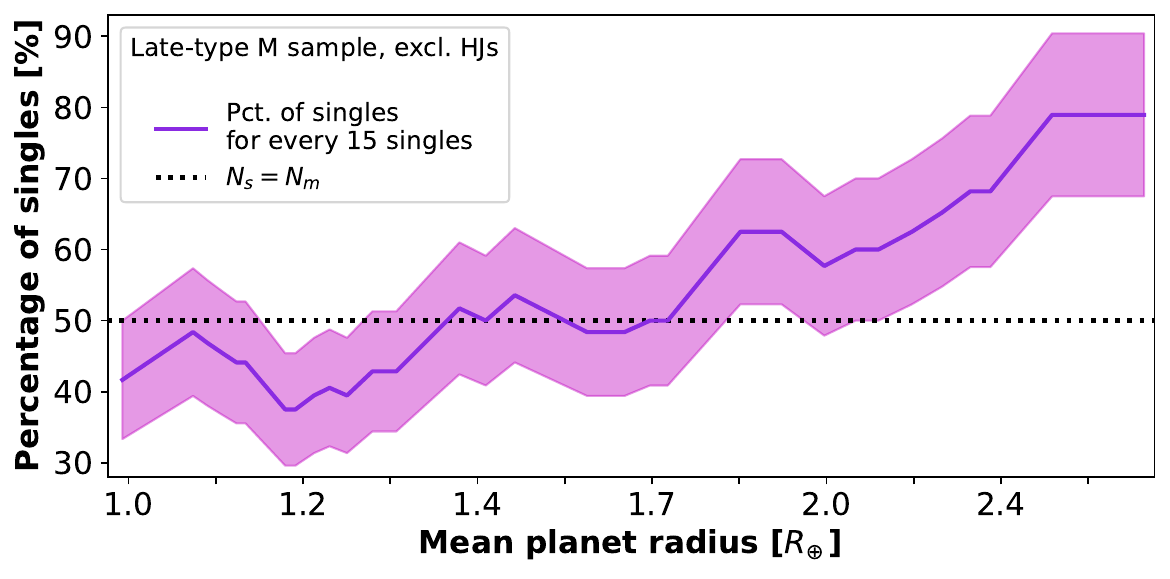}
        \caption{The late-type M sample, excluding the hot Jupiters.}
        \label{fig:pct_singles_c}
    \end{subfigure}
    \caption{Percentages (purple line) and binomial uncertainties (magenta region) of the planets that are singles as a function of planet radius (Sect.~\ref{sect:pct_singles}). These percentages are computed as a function of the mean of the planet radii in each consecutive bin comprising 100 singles in the FGK~sample (panel~a), 20 singles in the early-type M sample (panel~b), and 15 singles in the late-type M sample (panel~c). 
    The sharp decline for the FGK~sample occurs in the region of the multis overabundance at \mbox{$R \approx1.4-1.6\, R_\oplus$}.}
    \label{fig:pct_singles}
\end{figure}

In the early- and late-type M~samples, all the hot Jupiters are singles. In contrast to the FGK sample, even after excluding the hot Jupiters from both M-star samples, the singles and multis possess different distributions of planet types, as shown in Fig.~\ref{fig:pltypes_m}. 
Noticeably, these distributions are also different from those in the FGK sample as neither the singles nor the multis contain cold giants, and all the hot Jupiters, warm giants, and hot giants, except for one, reside in single-planet systems. 

\subsection{Percentages of singles and an overabundance of multis} \label{sect:pct_singles}

Excluding the hot Jupiters, we computed the fraction of singles for each planet type separately for the FGK and the two M-dwarf samples. The resulting fractions, along with their binomial uncertainties, are reported in Table~\ref{tab:pct_singles}. 
For instance, in the FGK sample, $\approx 95\%$ of the observed hot Jupiters, $\approx 61\%$ of the USPs, and $\approx 38\%$ of the cold giants belong to single-planet systems. 
Furthermore, we computed the fractional number of singles as a function of the planet radius for the FGK and the early- and late-type M samples, after removing the hot Jupiters, as plotted in Fig.~\ref{fig:pct_singles}.  
For this computation, we first sorted the planets in each of our three samples in order of increasing planet radius. Then, we applied a sliding-window method by calculating the percentage of planets that are singles in each moving bin containing 100 singles in the FGK sample, 20 singles in the early-type M sample, and 15 singles in the late-type M sample. In each bin, we included all the multis with radii between the minimum and maximum radius of the singles in that bin prior to computing the percentage as a function of the mean radius of the total planets in the bin. 

As shown in Fig.~\ref{fig:pct_singles_a}, in the FGK sample without hot Jupiters, the singles account for slightly less than 50\% of the planets with \mbox{$R \leq 6\, R_\oplus$} on average, regardless of the orbital period. However, the percentage function displays an unexpected and significant decline to $\approx 37\%$ at \mbox{$R\approx 1.5\, R_\oplus$}. 
Our analysis of this function showed that the percentage of singles exhibits a local minimum at \mbox{$R\approx 1.4-1.6\, R_\oplus$}. This implies that the fractional number of multis increases significantly within the specified radius range, which we hereafter refer to as the multis overabundance. 
In contrast to the FGK sample, the fraction of singles in neither the early- nor the late-M samples fluctuates, and it instead increases monotonically from $\approx40\%$ at a mean radius of $\approx1\, R_\oplus$ to more than 80\% at mean \mbox{$R\approx4\, R_\oplus$}. 

\begin{figure}
    \centering
    \begin{subfigure}{0.49\textwidth}
        \centering
        \includegraphics[width=\textwidth]{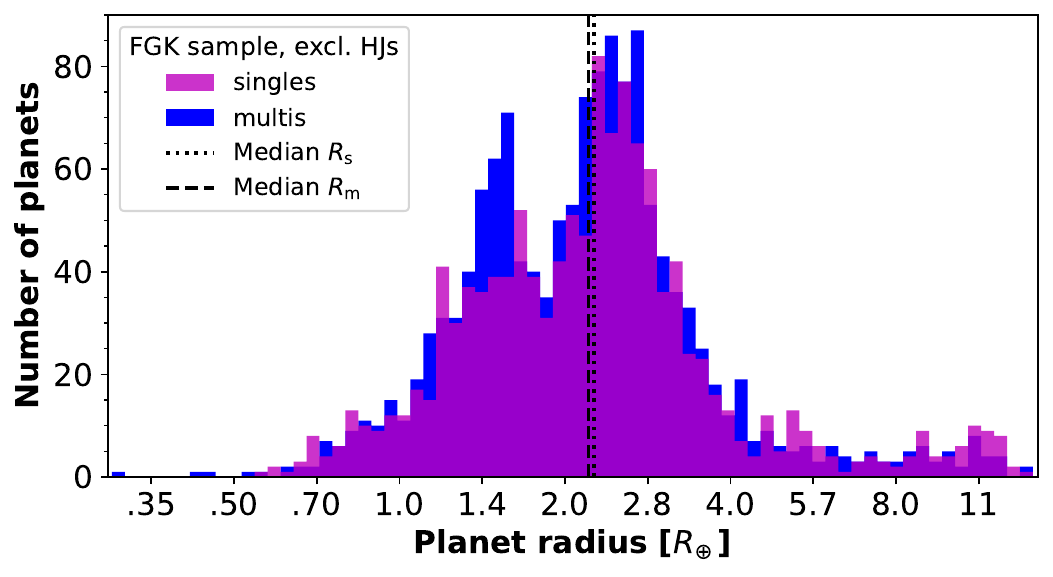}
        \caption{The FGK sample, after removing the hot Jupiters.}
        \label{fig:histograms_a}
    \end{subfigure}
    \par\medskip
    \begin{subfigure}{0.49\textwidth}
        \centering
        \includegraphics[width=\textwidth]{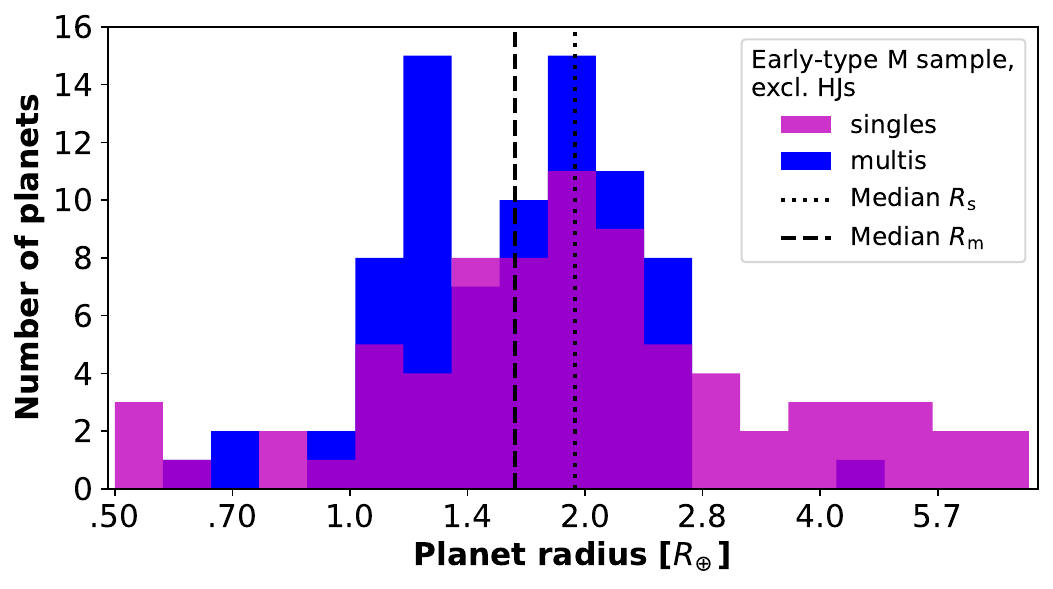}
        \caption{The early-type M sample, excluding the hot Jupiters.}
        \label{fig:histograms_b}
    \end{subfigure}
    \par\medskip
    \begin{subfigure}{0.49\textwidth}
        \centering
        \includegraphics[width=\textwidth]{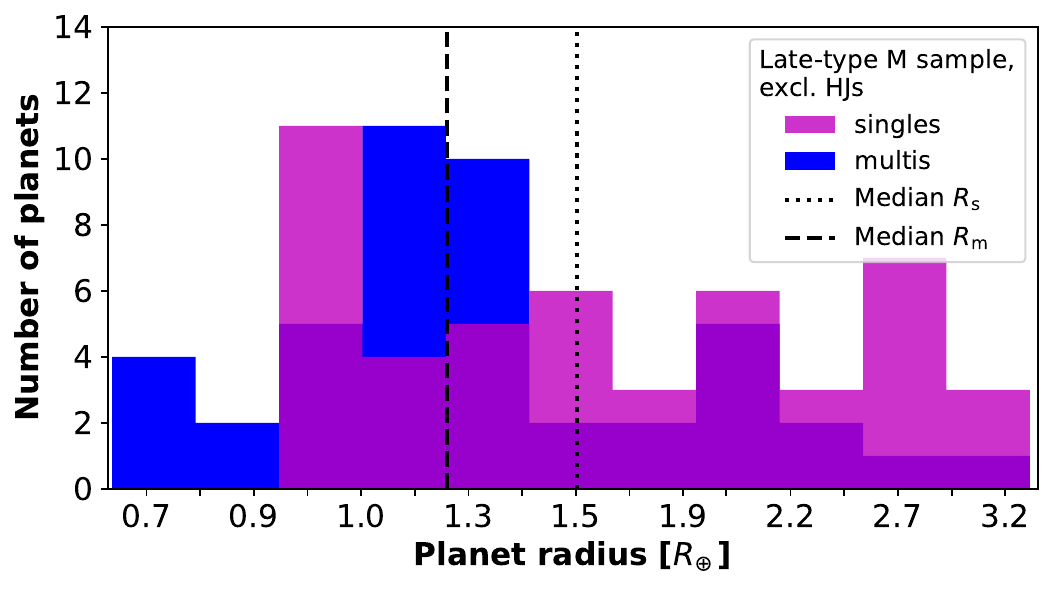}
        \caption{The late-type M sample, after removing the hot Jupiters.}
        \label{fig:histograms_c}
    \end{subfigure}
    \caption{Radius distributions of the singles (magenta) and multis (blue), with the median radii indicated by the vertical black lines. The singles and multis in the FGK sample (panel~a) have similar median radii and exhibit a dearth of planets at \mbox{$R \approx 1.8-1.9\, R_\oplus$}, which appears more pronounced for the multis because they are overabundant at \mbox{$R \approx 1.4-1.6\, R_\oplus$}. 
    In the early- and late-type M~samples, the singles have larger median radii than the multis.} 
    \label{fig:histograms}
\end{figure}

\begin{figure*}
\sidecaption
\centering
\includegraphics[width=12cm]{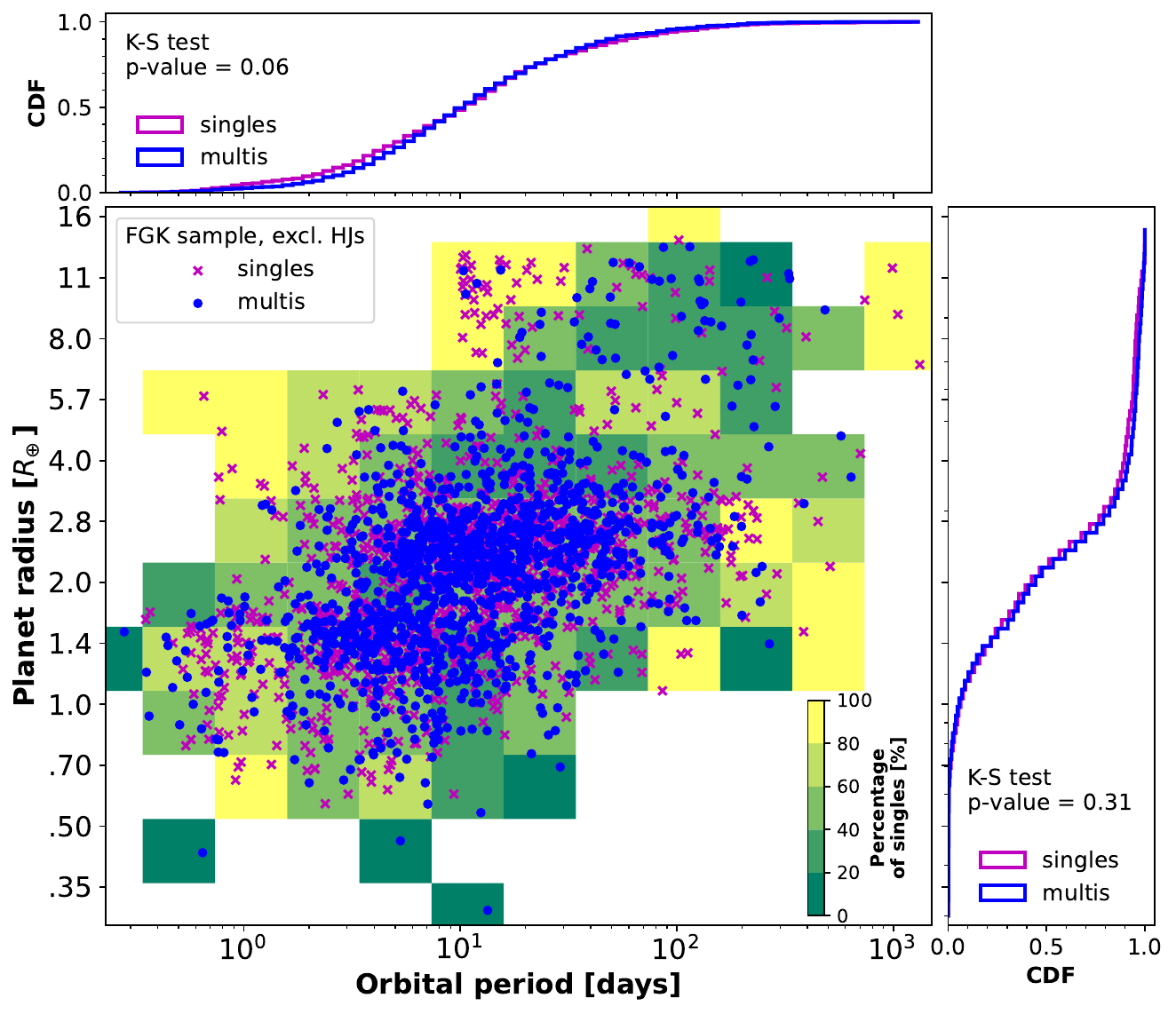}
\caption{Radii as a function of orbital periods of the singles (magenta crosses) and multis (blue bullets) in the FGK~sample, excluding hot Jupiters. The colour of each grid cell represents the percentage of planets that are singles within the cell, as indicated by the colour bar (see Table~\ref{tab:pct_singles}). The upper and the right panels display the empirical cumulative distribution functions of the orbital periods and planetary radii, respectively. The K-S test p-values indicate that the radii and orbital period  distributions of the singles and multis are statistically indistinguishable.}
\label{fig:mesh_fgk}
\end{figure*}

\begin{figure*}
\centering
    \begin{subfigure}[b]{0.49\textwidth}
        \centering
        \includegraphics[width=\textwidth]{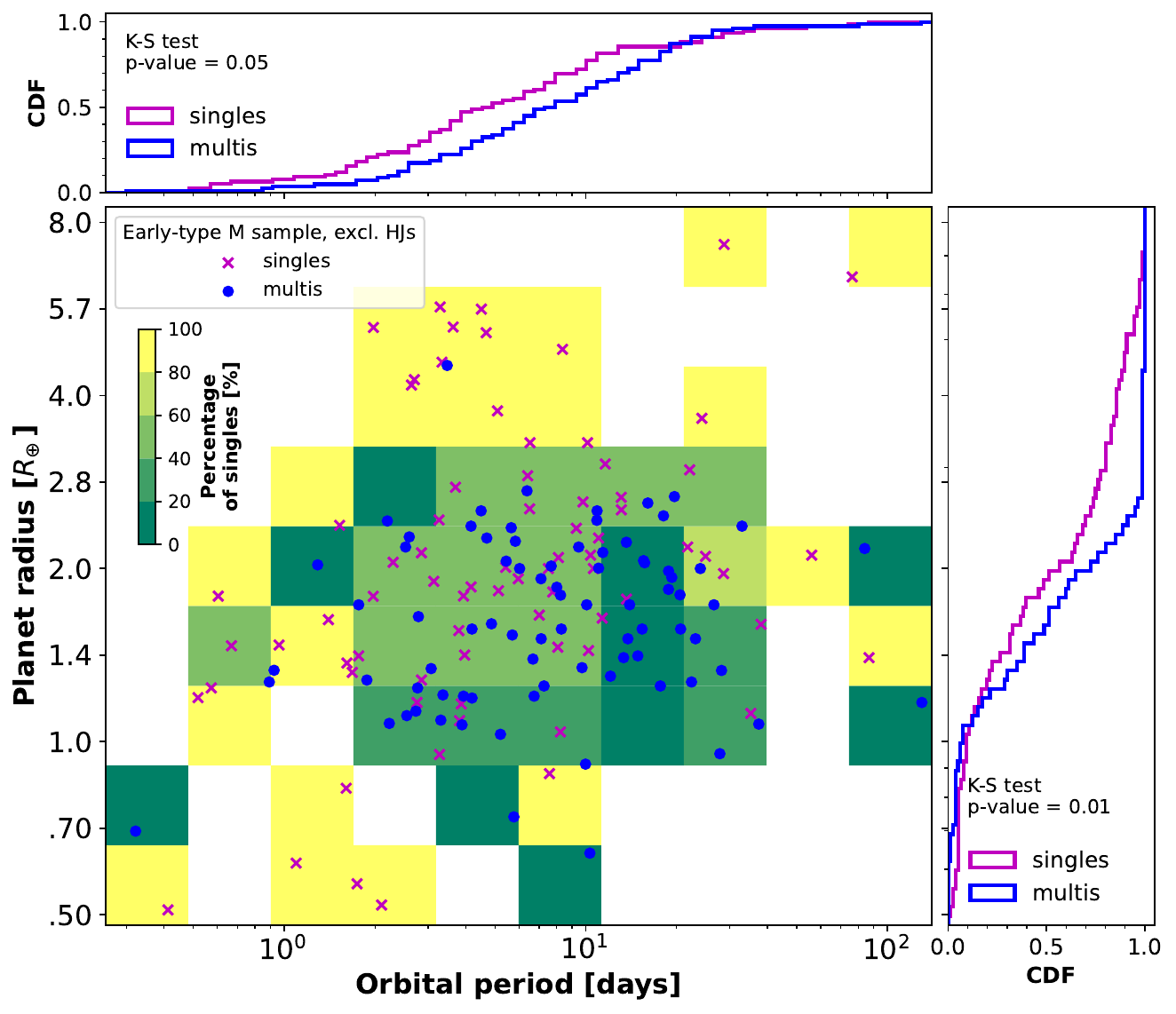}
        \caption{The early-type M sample, after removing the hot Jupiters.}
        \label{fig:mesh_m_a}
    \end{subfigure}
    %\par\medskip
    \begin{subfigure}[b]{0.49\textwidth}
        \centering
        \includegraphics[width=\textwidth]{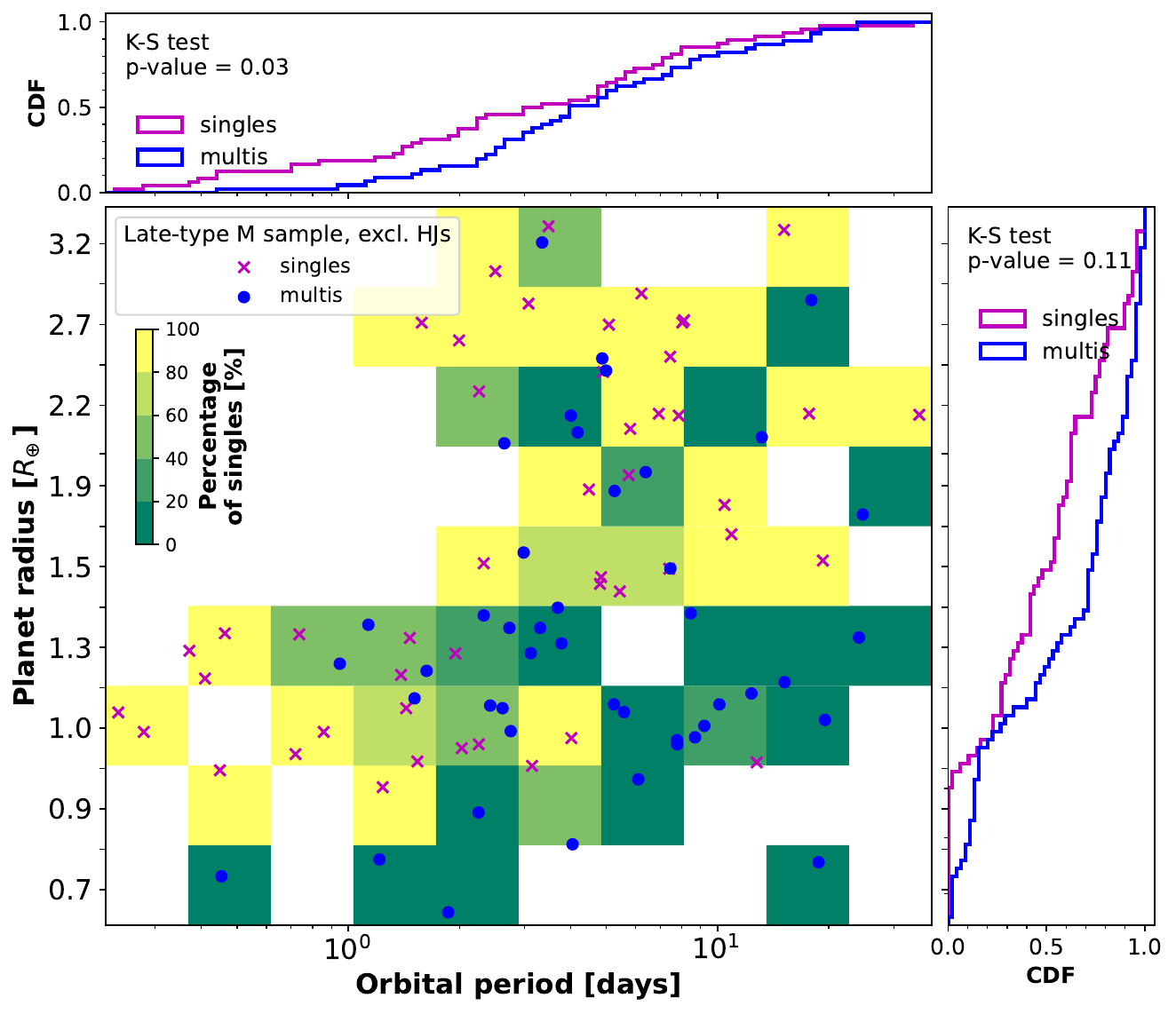}
        \caption{The late-type M sample, excluding the hot Jupiters.}
        \label{fig:mesh_m_b}
    \end{subfigure}
    \caption{Similar to Fig.~\ref{fig:mesh_fgk}, showing the singles and multis, excluding the hot Jupiters, in the period-radius plane, which is colour-coded by the percentage of singles. 
    %In both samples, particularly the late-type M sample (panel~b), singles are missing from the lower right-hand region of the graph, while they are more prevalent at large sizes and long orbital periods. 
    For both M~samples, the cumulative distribution functions indicate that both the radii and the orbital periods distributions of the singles and multis are different. These findings are uncertain due to the small sample sizes.}
    \label{fig:mesh_m}
\end{figure*}

\subsection{Distributions of planetary radii} \label{sect:radii}

The radius distributions of the multis and singles in the FGK and the two M~samples, after removing the hot Jupiters, are presented in Fig.~\ref{fig:histograms}. 
The singles and multis orbiting FGK stars display similar radius distributions, except for the region of the multis overabundance at \mbox{$R \approx 1.4 - 1.6\, R_\oplus$} where the number of multis increases significantly compared to the number of singles, as highlighted in Figs.~\ref{fig:pct_singles_a} and \ref{fig:histograms_a}. 
We found an indication of an overabundance of multis emerging for the planets orbiting M dwarfs as well, although at smaller radii of \mbox{$\approx 1.0 - 1.4\, R_\oplus$} (further discussed in Sect.~\ref{sect:overabundance}). We note, however, that this newly identified feature is uncertain given the low number of confirmed planets with M host stars. 

Furthermore, Fig.~\ref{fig:histograms_a} shows that both the multis and the singles hosted by FGK stars exhibit a deficit of planets at $\approx 1.8\, R_\oplus$ within the radius valley that separates super-Earths from sub-Neptunes \citep{fulton, van_eylen18}. This common feature is indicative of a universal mechanism producing the radius valley, which is independent of the planet multiplicity, consistent with earlier findings by \citet{weiss18b}. The multis, however, display a much deeper radius valley than the singles due to their overabundance at \mbox{$R\approx1.4-1.6\, R_\oplus$}. 

Figure \ref{fig:histograms} also indicates that the singles and multis orbiting M~dwarfs, especially those in the late-type M sample, exhibit different radius distributions. In the two M samples, the radius distributions of the singles are skewed towards higher values compared to those of the multis, and all the planets with \mbox{$R > 2.7\, R_\oplus$}, except for two multis, are singles. 
In contrast to \hyperref[ref:liberles]{Lib24}, who found that the multis around M and late-type K stars are larger than the singles on average, our analyses suggest the opposite: The singles are larger than the multis on average, both within the entire radius range and even only for the planets with \mbox{$R < 4\, R_\oplus$}. These results apply both to the M sample and to the planets orbiting M and late-type K stars with \mbox{$T_\mathrm{eff} < 4500$ K} selected from our catalogue after imposing the same temperature limit as \hyperref[ref:liberles]{Lib24}. 

We measured the significance of the differences between the radii of the singles and multis by conducting the following three comparison tests: a Kolmogorov–Smirnov (K-S) test, a two-sample Anderson-Darling (A-D) test, and a Mann–Whitney (M-W) U test (detailed in Sect.~\ref{apx:b}). The adopted null hypothesis was that the radii of the multis and singles can be drawn from the same distribution. 
We compared the radius distributions of the singles and multis in each of the FGK, early-, and late-type M~samples, and for the following three sub-samples: (i) all planets excluding hot Jupiters, (ii) planets with \mbox{$R < 4\, R_\oplus$}, and (iii) planets with \mbox{$R \geq 4\, R_\oplus$}.  Table~\ref{tab:pval} provides the numbers and the median radii of the planets, as well as the p-values from the three comparison tests for each sub-sample.

In order to account for the uncertainties in the radius measurements, we performed Monte Carlo simulations by drawing $10^3$ random samples with replacement, assuming a Gaussian distribution for the radius of each planet with its published value and errors corresponding to the median and $\pm1\sigma$ uncertainties of the distribution. 
In each of the $10^3$ runs, we computed the median radii and the statistics from the three comparison tests. The medians with the corresponding 16th and 84th percentiles of the resulting distributions for the computed test p-values and median radii are listed in Table~\ref{tab:pval}. 

Our findings indicate that the radii of the multis and singles in the FGK~sample without hot Jupiters are consistent with being drawn from the same distribution according to all three comparison tests (for example, a K-S p-value of $\approx 0.26$). 
As reported in Table~\ref{tab:pval}, the analyses also yield relatively high p-values for planets with \mbox{$R < 4\, R_\oplus$} (e.g. K-S p-\mbox{value $\approx 0.67$}) and lower p-values at \mbox{$R \geq 4\, R_\oplus$}, excluding hot Jupiters, (K-S p-\mbox{value $\approx 0.12$}). These results suggest that the singles and multis are consistent with originating from the same underlying population based on their radii, especially at radii \mbox{$< 4\, R_\oplus$}. 
This conclusion agrees with the data displayed in Fig.~\ref{fig:histograms_a}, showing that after removing the hot Jupiters, the multis and singles in the FGK~sample have similar median radii, particularly for planets with \mbox{$R < 4\, R_\oplus$}. Conversely, at \mbox{$R \geq 4\, R_\oplus$} the singles have a median \mbox{$R \approx 6.94\, R_\oplus$} and are therefore slightly larger than the multis, which have a median \mbox{$R \approx 6.10\, R_\oplus$}. 

In contrast to the FGK sample, the singles and multis in each of the two M~samples have different radius distributions with singles being larger on average, as shown in Fig.~\ref{fig:histograms} and Table~\ref{tab:pval}. 
Furthermore, when we excluded all the planets with \mbox{$R \geq 4\, R_\oplus$}, the radius distributions of the singles and multis orbiting early-type M dwarfs were statistically indistinguishable. We note that these results remain inconclusive due to the limited sample size of confirmed planets orbiting M dwarfs.

Figures \ref{fig:mesh_fgk} and \ref{fig:mesh_m} display the radii as a function of the orbital periods of the singles and multis, excluding the hot Jupiters, in the FGK, early-, and late-type M samples. The percentage of the detected planets that are singles in each grid cell of the graph is also indicated. 
The cumulative distribution functions shown in each of these figures help us to compare the distributions of the radii and orbital periods of the singles with those of the multis within each of the three samples. 
As seen in Fig.~\ref{fig:mesh_fgk}, the singles and multis in the FGK sample, after removing the hot Jupiters, have similar cumulative distributions of the radii and orbital periods. Notably, there is no significant difference in the period-radius planes of the multis and singles.  
In contrast, Fig.~\ref{fig:mesh_m} shows that the singles and multis in each of the two M~samples do not display similar cumulative distribution functions. 

Figure \ref{fig:mesh_m} also indicates that the planets orbiting early- and late-type M dwarfs have shorter orbital periods on average and are smaller and less diverse than the planets hosted by FGK~stars. 
Particularly, except for two multis, all the planets with \mbox{$R > 2.7\, R_\oplus$} in the two M samples are singles, of which only two singles in the early-type M sample have \mbox{$P > 25$ days}. In contrast, in the FGK sample, multis account for $\approx59\%$ of the planets with \mbox{$P > 25$ days} and \mbox{$R > 3\, R_\oplus$}, even though the five planets with the longest orbital periods are singles. 

\begin{table*}
\begin{center}
\caption{Properties of the singles and multis in the FGK, early-, and late-type M samples.}
\label{tab:pval}
\begin{tabular}{llccccccccc}
\hline\hline
\noalign{\smallskip}
Spec. & Sub-sample & $N_s$ & $N_m$ & $\widetilde{R_s}$ & $\widetilde{R_m}$ & K-S & A-D & M-W\\
type & & & & [$R_\oplus$] & [$R_\oplus$] & p-value & p-value & p-value\\
\noalign{\smallskip}
\hline
\noalign{\smallskip}
\multirow{5}{*}{FGK} & Full~samp. & 1587 & 1397 &  2.57±0.02 & 2.21±0.01 & $<10^{-14}$ & $\leq10^{-3}$ & $<10^{-26}$\\
& Excl.~HJs\tablefootmark{a} & 1232 & 1379 & 2.24±0.01 & 2.19±0.01 & 0.26±0.12 & 0.11±0.04 & 0.21±0.07\\
& \mbox{$R < 4\, R_\oplus$} & 1099 & 1258 & 2.10±0.01 & 2.09±0.01 & 0.67±0.17 & $\geq0.25$ & 0.74±0.15\\
& \mbox{$R \geq 4\, R_\oplus$}\tablefootmark{b} & 133 & 121 & 6.94±0.20 & 6.10±0.09 & 0.12±0.06 & 0.04±0.03 & 0.04±0.02\\
\hline
\noalign{\smallskip}
\multirow{3}{*}{M0-2} & Full~samp. & 87 & 80 &  2.08±0.04 & 1.64±0.04 & $<10^{-4}$ & $\leq10^{-3}$ & $<10^{-3}$\\
& Excl.~HJs\tablefootmark{a} & 76 & 80 & 1.94±0.04 & 1.64±0.04 & 0.012±0.009 & 0.004±0.002 & 0.016±0.009\\
& $R < 4\, R_\oplus$ & 65 & 79 & 1.79±0.04 & 1.62±0.04 & 0.37±0.13 & 0.15±0.05 & 0.27±0.09\\
\hline
\noalign{\smallskip}
\multirow{3}{*}{M3-9} & Full~samp. & 56 & 45 & 1.81±0.07 & 1.22±0.03 & 0.003±0.003 & $\leq10^{-3}$ & $<10^{-3}$\\
& Excl.~HJs\tablefootmark{a} & 48 & 45 & 1.53±0.05 & 1.22±0.03 & 0.047±0.029 & 0.017±0.009 & 0.018±0.10\\
& $R < 4\, R_\oplus$ & 48 & 45 & 1.53±0.05 & 1.22±0.03 & 0.047±0.029 & 0.017±0.009 & 0.018±0.10\\
\hline
\end{tabular}
\end{center}
\tablefoot{$s$ and $m$ stand for singles and multis, respectively, while $N$ and $\widetilde{R}$ represent the number and the median radii of the planets in each sample, respectively. 
Spectral types M0-2 and M3-9 correspond to early- and late-type M dwarfs, respectively. 
The p-values indicate the significance of the differences between $R_s$ and $R_m$ in each sample, as measured with three tests: the Kolmogorov–Smirnov (K-S), the two-sample Anderson-Darling (A-D), and the Mann–Whitney (M-W) U test. All the reported $\pm1\sigma$-uncertainties are computed via Monte Carlo simulations (Sect.~\ref{sect:radii}).}
\tablefoottext{a}{This sub-sample comprises the singles and multis in the sample, after removing the hot Jupiters.} 
\tablefoottext{b}{This sub-sample contains the giants with $R \geq 4\, R_\oplus$, excluding the hot Jupiters.}
\end{table*}

\subsection{Subsets of \textit{Kepler} planets} \label{sect:subsets}

In our initial catalogue presented in Sect.~\ref{sect:samples}, we included nearly all the confirmed transiting planets in order to acquire a large sample of both singles and multis. This sample is diverse and heterogeneous as it contains various planets detected primarily by \textit{Kepler}, K2, and TESS. 
In order to evaluate our results reported in Table~\ref{tab:pval}, we compiled three additional smaller and more homogeneous subsets consisting only of confirmed \textit{Kepler} planets. 
All three subsets were constructed using the same planets to create a consistent basis for comparison and to evaluate how the variations in references, measured values, and degree of sample homogeneity affect the statistical outcomes.

Two of the new subsets were selected from the planet catalogue provided by \citet[][hereafter \color{blue}L24]{lissauer24}, which contains stellar and planetary parameters for more than 4300~transiting \textit{Kepler} Objects of Interest (KOI) candidates and confirmed planets. 
The third subset comprises the same \textit{Kepler}~planets as the first two subsets but with parameters from our initial catalogue, which contains only confirmed planets. 
We cross-matched the \hyperref[ref:lissauer24]{L24}~catalogue with our catalogue without applying any upper limit on the radius uncertainties in order to select all the planets present in both datasets. 
This process resulted in 1080~singles and 1055~multis orbiting FGK~stars, as well as 24~singles and 42~multis hosted by M~dwarfs, after adopting the effective temperatures in our catalogue.\footnote{There are 21 singles and 49 multis in our initial catalogue that are not present in the \hyperref[ref:lissauer24]{L24}~catalogue. These planets were therefore not included in any of the three new subsets.} 

Since the numbers of singles and multis hosted by M~dwarfs are small and highly unequal, any analysis of these planets would likely be biased and unreliable (further discussed in Sect.~\ref{sect:inconsistency}). 
Therefore, we compiled the following three subsets, each containing the same 2135 cross-matched planets orbiting FGK~stars, but with planetary parameters selected from different sources: 
\begin{enumerate}[label=\Roman*)]
  \setlength\itemsep{1.5mm}
    \item the \hyperref[ref:lissauer24]{L24} subset~I (based on the first set of planetary parameters from \hyperref[ref:lissauer24]{L24}, which prioritizes accuracy),
    \item the \hyperref[ref:lissauer24]{L24} subset~II (based on the second parameter set from \hyperref[ref:lissauer24]{L24}, which has been derived in a homogeneous manner),
    \item this work's subset III (containing the corresponding parameters from our initial planet catalogue).
\end{enumerate}

We conducted similar comparison tests as in Sect.~\ref{sect:radii} by first dividing each of the three subsets into distinct sub-samples based on the planet types, prior to comparing the singles and multis. The numbers and median radii of the planets in these sub-samples as well as the p-values from the K-S, A-D, and M-W U tests (following $10^3$~Monte Carlo simulations) are reported in Table~\ref{tab:pval_kepler}. The results are consistent for all three subsets, indicating that the radius distributions of the singles and multis are statistically indistinguishable for the full subsets with and without hot Jupiters and for the sub-samples of planets with \mbox{$R < 4\, R_\oplus$}. 

The p-values from the comparison tests performed on the FGK~sample from our initial catalogue (Table~\ref{tab:pval}) agree overall with the p-values listed in Table~\ref{tab:pval_kepler}. The only difference arises when hot Jupiters are included: In contrast to our initial full FGK~sample, the singles and multis in all three full subsets in Table~\ref{tab:pval_kepler} are statistically indistinguishable based on the K-S and the M-W U tests. 
Additionally, as reported in Table~\ref{tab:pval} for our FGK sample after removing the hot Jupiters, the singles exhibit a larger median radius than the multis for planets with \mbox{$R \geq 4\, R_\oplus$}. This trend appears strengthened in the three new subsets at \mbox{$R \geq 4\, R_\oplus$}, excluding hot Jupiters, particularly in the homogeneous \hyperref[ref:lissauer24]{L24} subset~II. 

In conclusion, there are only minor differences between the three subsets in Table~\ref{tab:pval_kepler}, based on the analysis we performed. The \hyperref[ref:lissauer24]{L24}~subset~I, the uniformly derived \hyperref[ref:lissauer24]{L24}~subset~II, and our work's subset~III display similar radius distributions and show consistent results, regardless of the chosen references for the planetary radii. 
This further implies that the main distinction between the results for our initial FGK~sample and those for the three subsets is driven by the inclusion of the non-KOI transiting planets in our initial sample and by the correspondingly increased sample size.

\subsection{Inconsistency with a previous study} \label{sect:inconsistency}

In a recent study, \hyperref[ref:liberles]{Lib24} analysed a sample of KOIs comprising a total of 149~singles and 141~multis orbiting M and late-type K~stars with \mbox{$T_\mathrm{eff} < 4500$ K}. 
For the entire sample and only for planets with \mbox{$R < 6\, R_\oplus$}, the authors identified that the radii of the singles and multis cannot be drawn from the same distribution, and that the multis are larger than the singles on average. 
In contrast, our analyses indicate that the multis in our early- and late-type M~samples are smaller than the singles, as shown in Table~\ref{tab:pval} and Fig.~\ref{fig:histograms}. 
However, these findings must be interpreted with caution given the limited sample sizes studied in both works. 

As shown in Fig.~\ref{fig:boxplots} and Table~\ref{tab:pval}, the planets, especially the multis, orbiting late-type M~dwarfs have smaller radii overall than the planets hosted by FGK stars and early-type M~dwarfs. 
Additionally, in contrast to the late-M sample, the radius distributions of the singles and multis with \mbox{$R < 4\, R_\oplus$} orbiting early-type M~dwarfs were found to be statistically indistinguishable (K-S \mbox{p-value $\approx0.37$}). 
Therefore, planets orbiting late-M, early-M, and FGK~stars should be analysed as separate samples (see e.g. Tables~\ref{tab:pct_pltypes}–\ref{tab:pval}) in order to robustly test the differences between the singles and multis within and across the distinct samples. 

Furthermore, we cross-matched the planet sample examined by \hyperref[ref:liberles]{Lib24} with the list of KOIs from the NASA Exoplanet Archive \citep{christiansen} and found that the sample contains 149~singles (68 candidates and 81 confirmed) and 141 multis (12 candidates and 129 confirmed). 
Adopting the planetary radii used by \hyperref[ref:liberles]{Lib24}, we compared the singles and multis in their sample and obtained results that agree with those reported by the authors. 
Additionally, we removed all the planet candidates from the \hyperref[ref:liberles]{Lib24}~sample and only compared the confirmed KOIs. Our analyses showed that the radius distributions of all the confirmed singles and multis in this dataset as well as only of those with \mbox{$R < 6\, R_\oplus$} exhibit no significant differences. This highlights the discrepancy between the authors' initial sample and the set that only contains confirmed planets. 

In order to further compare our results with those reported by \hyperref[ref:liberles]{Lib24}, we compiled two new samples after implementing the same upper limit on the host star temperature as was applied by the aforementioned authors. 
The first new sample comprised all the planets orbiting stars with \mbox{$T_\mathrm{eff}<4500$ K} selected from our initial catalogue without applying any upper limit on the radius uncertainty. This amounted to 268~singles and 257~multis, 107, 180, and 238~planets of which have late-M, early-M, and late-K~host stars, respectively. This new sample thus contains three times as many singles and twice as many multis as those that were confirmed in the \hyperref[ref:liberles]{Lib24} dataset.
For this new sample as well as only for the subset of planets with \mbox{$R < 6\, R_\oplus$}, we found significant evidence that the radius distributions of the singles and multis are different, and that the median radius of the singles is larger. 
However, after we removed the planets hosted by late-type M dwarfs from this sample, our analyses revealed that the radii of the remaining multis and singles with \mbox{$R < 5\, R_\oplus$} appear to be statistically indistinguishable (K-S \mbox{p-value $\approx0.12$}). This result emphasises the importance of accounting for the planets orbiting late-type M dwarfs, since they are smaller on average than the rest of the planets, as shown for instance in Fig.~\ref{fig:boxplots}. 

The second new sample used for comparisons between \hyperref[ref:liberles]{Lib24}'s and our results contains all the confirmed KOIs orbiting stars with \mbox{$T_\mathrm{eff} < 4500$ K} selected from the \hyperref[ref:lissauer24]{L24} catalogue (described in Sect.~\ref{sect:subsets}). We adopted the first set of parameters provided in their catalogue since not all the host stars had a temperature value listed in the second set of parameters. 
Our analyses identified the radius distributions of the resulting 96~singles and 124~multis to be statistically indistinguishable for the entire dataset as well as only for the planets with \mbox{$R < 6\, R_\oplus$}, thereby contradicting \hyperref[ref:liberles]{Lib24}'s finding. 
Since this sample only contains nine planets hosted by late-type M dwarfs, these results mainly reflect the planet population around early-M and late-K~stars.

\section{Discussion} \label{sect:discussion}
\subsection{The multis overabundance} \label{sect:overabundance}

We investigated the planet multiplicity distributions in several samples of confirmed transiting planets and compared the planet type distributions, the fractional numbers, and the radii of the singles with those of the multis in different host star classes. These analyses are essential for determining whether the observed singles are genuine singles or members of multi-planetary systems in which additional planets have not yet been detected. 

As shown in Figs.~\ref{fig:pct_singles_a} and \ref{fig:histograms_a} for the FGK sample, there is an overabundance of multis (184) compared to singles (109) at radii of \mbox{$\approx 1.4-1.6\, R_\oplus$}. When examining only the \textit{Kepler}~planets from this sample, we identified an overabundance of multis within the same radius interval of \mbox{$R\approx 1.4-1.6\, R_\oplus$}, thereby indicating that the multis overabundance does not emerge from our catalogue's combination of planets detected by different telescopes. 
Correspondingly for the early- and late-type M samples, Fig.~\ref{fig:histograms} indicates that the multis are more prevalent than the singles at \mbox{$R\approx 1.0-1.4\, R_\oplus$}. This finding remains tentative, however, owing the insufficiently large numbers of confirmed single- and multi-planetary systems hosted by M~dwarfs.

We investigated whether the planets within the radius range of the multis overabundance in the FGK~sample have distinctive or atypical properties compared to the rest of the singles and multis, which follow a similar distribution everywhere outside the mentioned region, as seen in Fig.~\ref{fig:histograms_a}. 
The higher occurrence of multis relative to singles at \mbox{$R\approx 1.4-1.6\, R_\oplus$} might indicate either that planets of these sizes are intrinsically more likely to reside in multi-planet systems, or that observational biases favour their detection in systems hosting multiple planets, for instance due to low mutual inclinations. 

A noticeable feature for the early- and late-type M samples in Fig.~\ref{fig:mesh_m} is that the singles and multis in the region of the multis overabundance at \mbox{$R\approx 1.0-1.4\, R_\oplus$} do not have similar orbital periods. Nearly all the singles with radii in this range possess short orbital periods of \mbox{$P < 2$ days}, while the vast majority of the multis have \mbox{$P > 2$ days}. 
This feature might result from statistical biases due to the small sample size, and the discovery of additional planets is therefore required in order to verify this trend. 
Conversely, the planets in the radius range of the multis overabundance orbiting FGK~stars possess similar orbital periods, as observed in Fig.~\ref{fig:mesh_fgk}.

Furthermore, there are only 26 pairs of adjacent planets in the FGK~sample in which both planets possess radii of \mbox{$\approx 1.4-1.6\, R_\oplus$}, while the remaining planets in this radius interval have adjacent neighbours with smaller and/or larger radii. Therefore, the multis overabundance contains 184 multis in 158 different planetary systems, thus indicating that the overabundance of multis is not caused by an abundance of systems that contain peas-in-a-pod, similarly sized planets with \mbox{$R\approx1.4-1.6\, R_\oplus$}. 
Consequently, the multis overabundance warrants further investigations in order to determine its probable origin. 

\begin{figure}
\resizebox{\hsize}{!}{\includegraphics{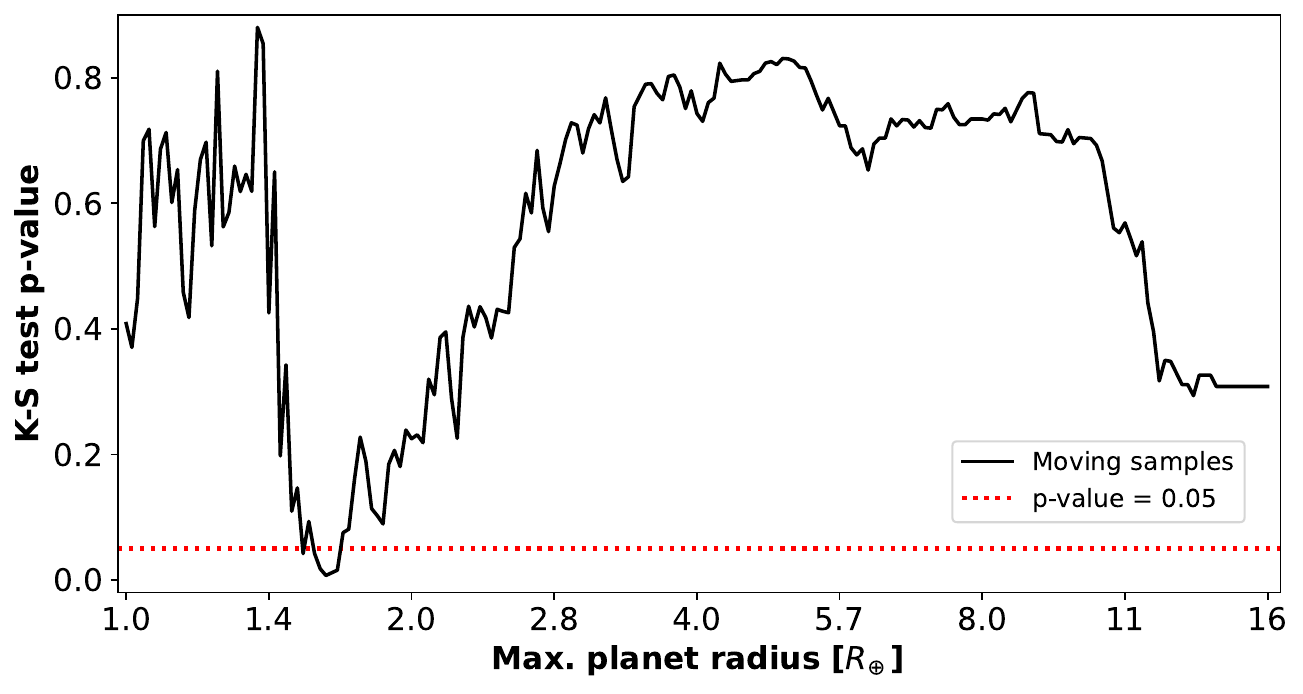}}
\caption{K-S test p-values for the singles and multis in the FGK~sample, excluding hot Jupiters, computed as a function of the maximum radius. Each consecutive test compares only the planets with radii smaller than the radius indicated on the x-axis. The multis and singles are significantly different when the planets within the region of the multis overabundance at \mbox{$R \approx 1.4-1.6\, R_\oplus$} are included in the test. Only after including larger planets do the p-values increase again above the 0.05 significance level.}
\label{fig:pv_maxrad}
\end{figure}

In addition to the tests described in Sect.~\ref{sect:radii}, we computed the K-S, the A-D, and the M-W U tests on the multis and singles, excluding hot Jupiters, in the FGK sample as a function of increasing maximum radius $R_\mathrm{max}$. In each test, we only included the planets with radii smaller than a specified value, incrementally increasing from 1 to $16\, R_\oplus$. The K-S p-values from this analysis are plotted in Fig.~\ref{fig:pv_maxrad}, indicating that the radii of the multis and singles are not significantly different overall. 
However, the p-values and, thus, the similarities between the singles and multis abruptly decrease at \mbox{$R_\mathrm{max}\approx 1.4\, R_\oplus$} and remain below the 0.05 significance level until \mbox{$R_\mathrm{max}\approx 1.6\, R_\oplus$}. 
This feature is likely associated with the overabundance of multis, since the p-values decline sharply when planets with \mbox{$R \approx 1.4\, R_\oplus$} are included in the analysis. Only after leaving the regime of the multis overabundance at \mbox{$R > 1.6\, R_\oplus$} do the p-values begin to increase, indicating that the radius distributions of singles and multis again become statistically indistinguishable. 

In our previous study of confirmed multi-planetary systems, we identified a correlation between the masses and orbital spacings of adjacent planets, as well as enhanced orbital spacing dispersions of inner planets in systems that also host outer giant planets. 
These properties suggest that large outer planets might affect the inner planets and their orbital configurations dynamically, potentially exciting their mutual inclinations, and thereby, reducing the number of inner planets that are detectable via transits \citep{lai}. 
This hypothesis might explain the statistical similarity between the observed singles and multis (excluding hot Jupiters) if a fraction of the apparent singles actually reside in multi-planetary systems with planets that remain undetected due to orbital misalignment. 
Additionally, there are other possible mechanisms that can elevate the mutual inclinations in a planetary system during its formation and evolution, such as planet–planet scattering \citep{weiss18b}, planet interactions and breaking of resonant chains \citep{izidoro}, and spin-orbit misalignment \citep{spalding}. 
These processes are also likely to increase the orbital eccentricities, thus explaining why the observed singles possess higher eccentricities on average than the multis \citep{xie, lissauer24}. 
Nevertheless, it has not yet been studied whether these mechanisms are also able to predict the multis overabundance found in this work. 

\subsection{Observational biases}

\citet{weiss18b}, \citet{berger}, and \citet{zhu} found no significant differences in the distributions of the stellar mass, radius, age, metallicity, temperature, or projected rotational velocity between stars hosting multiple planets and those hosting singles. Therefore, we considered in our analyses that the selection and observational biases associated with the host stars are the same for the singles and multis, hence not affecting the comparisons between them. 
We investigated the possibility that a fraction of the detected single-planet systems might in fact host additional as yet undetected planets and should therefore be reclassified as multis. By contrast, we assumed that all the multis in our catalogue are genuine, owing to their high reliability and the significant probability that multiple planets occur within the same system \citep{lissauer14}. 

For the entire FGK~sample and each of its three sub-samples listed in Table~\ref{tab:pval}, we conducted a bootstrap analysis in which one-third of the singles were randomly selected and assigned as multis. We iterated this process $10^3$~times, computing the K-S, A-D, and M-W U tests between the new singles and multis in each run. 
The medians of the p-value distributions resulting from this analysis agree with the values for the initial samples listed in Table~\ref{tab:pval}. They indicate that the radii of the resampled multis and singles in each of the three FGK sub-samples without hot Jupiters can be drawn from the same distribution. 
Importantly, these results suggest that the observed similarity between the singles and multis is likely to persist and might even increase as additional adjacent planets or outer giants are discovered in already known systems \citep[e.g.][]{persson, lillo-box, orell-miquel}.

Complementarily, we considered the possibility that a fraction of the singles in our catalogue are in fact false positives. In a process similar to the one described above, we performed $10^3$ iterations on the FGK~sample without hot Jupiters and randomly selected one-third of the singles that lacked radial-velocity measurements. These planets were discarded from the sample in each iteration prior to computing the K-S, A-D, and \mbox{M-W U} tests between the multis and the resampled singles. 
The medians of the resulting p-values are consistent with the p-values for the initial samples reported in Table~\ref{tab:pval}. This indicates that the radius distributions of the singles and multis, excluding hot Jupiters, in each of the three FGK sub-samples are statistically indistinguishable even after demoting some of the singles to false positives. 
Notably, the distribution of the false-positive probabilities of the confirmed KOIs has been shown to display a low median of $<1\%$ \citep{morton}. 
Given that approximately 75\% of the singles in our catalogue are confirmed KOIs, our choice of discarding one-third of the singles likely represents an overestimated large number. Removing a smaller proportion of singles would therefore be expected to have an even more negligible effect on the initial results reported in Table~\ref{tab:pval}. 

\newpage
\section{Conclusions} \label{sect:conclusion}

We compared the planet types, the fractional numbers, and the radius distributions of the confirmed transiting singles and multis in different host star classes divided into the FGK-, the early \mbox{M-,} and the late M-star samples. The main conclusions are listed below:\\[2mm]
The mean planet radii of both the multis and the singles increase with the host star temperature.\\[2.5mm]
In the full FGK~sample including hot Jupiters, the singles are larger on average than the multis.\\[2.5mm]
For the FGK sample, we found the following after removing the hot Jupiters:
\begin{itemize}
  \setlength\itemsep{2.2mm}
    \item The planet types and their fractional numbers comprising the singles and multis are similar.
    \item The radius distributions of the singles and multis are statistically indistinguishable overall and display similar median radii, particularly at \mbox{$R < 4\, R_\oplus$}. 
    \item We identified an abundance of multis, compared to singles, at \mbox{$R \approx 1.4-1.6\, R_\oplus$}, which we referred to as the multis overabundance. 
\end{itemize}
\vspace{0.6mm}
For the early- and late-type M samples, the following findings are inconclusive because the sample sizes were small:
\begin{itemize}
  \setlength\itemsep{2.2mm}
    \item Both the multis and the singles orbiting late-type M dwarfs are smaller than those with early-type M hosts. 
    \item The radii of the singles and multis in neither the early- nor the late-type M samples can be drawn from the same distribution. Only the radius distributions of the planets with \mbox{$R < 4\, R_\oplus$} orbiting early-M stars are statistically indistinguishable.
    \item The singles are larger on average than the multis in the late-type M sample.
    \item We found an overabundance of multis, compared to singles, at \mbox{$R\approx 1.0-1.4\, R_\oplus$}.
\end{itemize}

\vspace{5mm}
\begin{acknowledgements}
We thank the anonymous referee for the insightful comments that helped to improve this manuscript. This research has made use of the NASA Exoplanet Archive, which is operated by the California Institute of Technology, under contract with the National Aeronautics and Space Administration under the Exoplanet Exploration Program. A.M. and C.M.P. acknowledge the generous support of GENIE at Chalmers.
\end{acknowledgements}

\bibliographystyle{aa}
\bibliography{bibliography}

@ARTICLE{armitage,
       author = {{Armitage}, Philip J.},
        title = "{Planet formation theory: an overview}",
      journal = {arXiv e-prints},
         year = 2024,
        month = dec,
          eid = {arXiv:2412.11064},
        pages = {arXiv:2412.11064},
          doi = {10.48550/arXiv.2412.11064},
archivePrefix = {arXiv},
       eprint = {2412.11064},
 primaryClass = {astro-ph.EP},
       adsurl = {https://ui.adsabs.harvard.edu/abs/2024arXiv241211064A}
}

@ARTICLE{ballard,
       author = {{Ballard}, Sarah and {Johnson}, John Asher},
        title = "{The Kepler Dichotomy among the M Dwarfs: Half of Systems Contain Five or More Coplanar Planets}",
      journal = {\apj},
         year = 2016,
        month = jan,
       volume = {816},
       number = {2},
          eid = {66},
        pages = {66},
          doi = {10.3847/0004-637X/816/2/66},
archivePrefix = {arXiv},
       eprint = {1410.4192},
 primaryClass = {astro-ph.EP},
       adsurl = {https://ui.adsabs.harvard.edu/abs/2016ApJ...816...66B}
}

@ARTICLE{berger,
       author = {{Berger}, T. and {Huber}, D. and {Gaidos}, E. and {van Saders}, J. and {Weiss}, L.},
        title = "{The Gaia-Kepler Stellar Properties Catalog. II. Planet Radius Demographics as a Function of Stellar Mass and Age}",
      journal = {\aj},
         year = 2020,
        month = sep,
       volume = {160},
       number = {3},
          eid = {108},
        pages = {108},
          doi = {10.3847/1538-3881/aba18a},
archivePrefix = {arXiv},
       eprint = {2005.14671},
 primaryClass = {astro-ph.EP},
       adsurl = {https://ui.adsabs.harvard.edu/abs/2020AJ....160..108B}
}

@INCOLLECTION{burn,
       author = {{Burn}, Remo and {Mordasini}, Christoph},
        title = {Planetary population synthesis},
    booktitle = {Handbook of Exoplanets},
         year = 2024,
    publisher = {Springer},
          eid = {143-2},
        pages = {143-2},
          doi = {10.1007/978-3-319-30648-3_143-2},
       adsurl = {https://ui.adsabs.harvard.edu/abs/2024haex.book..143B}
}

@ARTICLE{christiansen,
       author = {{Christiansen}, Jessie L. and {McElroy}, Douglas L. and {Harbut}, Marcy and {Ciardi}, David R. and {Crane}, Megan and {Good}, John and {Hardegree-Ullman}, Kevin K. and {Kesseli}, Aurora Y. and {Lund}, Michael B. and {Lynn}, Meca and {Muthiar}, Ananda and {Nilsson}, Ricky and {Oluyide}, Toba and {Papin}, Michael and {Rivera}, Amalia and {Swain}, Melanie and {Susemiehl}, Nicholas D. and {Tam}, Raymond and {van Eyken}, Julian and {Beichman}, Charles},
        title = "{The NASA Exoplanet Archive and Exoplanet Follow-up Observing Program: Data, Tools, and Usage}",
      journal = {The Planetary Science Journal},
         year = 2025,
        month = aug,
       volume = {6},
       number = {8},
          eid = {186},
        pages = {186},
          doi = {10.3847/PSJ/ade3c2},
archivePrefix = {arXiv},
       eprint = {2506.03299},
 primaryClass = {astro-ph.EP},
       adsurl = {https://ui.adsabs.harvard.edu/abs/2025PSJ.....6..186C}
}

@ARTICLE{fulton,
       author = {{Fulton}, Benjamin J. and {Petigura}, Erik A. and {Howard}, Andrew W. and {Isaacson}, Howard and {Marcy}, Geoffrey W. and {Cargile}, Phillip A. and {Hebb}, Leslie and {Weiss}, Lauren M. and {Johnson}, John Asher and {Morton}, Timothy D. and {Sinukoff}, Evan and {Crossfield}, Ian J.~M. and {Hirsch}, Lea A.},
        title = "{The California-Kepler Survey. III. A Gap in the Radius Distribution of Small Planets}",
      journal = {\aj},
         year = 2017,
        month = sep,
       volume = {154},
       number = {3},
          eid = {109},
        pages = {109},
          doi = {10.3847/1538-3881/aa80eb},
archivePrefix = {arXiv},
       eprint = {1703.10375},
 primaryClass = {astro-ph.EP},
       adsurl = {https://ui.adsabs.harvard.edu/abs/2017AJ....154..109F}
}

@ARTICLE{gaidos,
       author = {{Gaidos}, E. and {Mann}, A.~W. and {Kraus}, A.~L. and {Ireland}, M.},
        title = "{They are small worlds after all: revised properties of Kepler M dwarf stars and their planets}",
      journal = {\mnras},
         year = 2016,
        month = apr,
       volume = {457},
       number = {3},
        pages = {2877-2899},
          doi = {10.1093/mnras/stw097},
archivePrefix = {arXiv},
       eprint = {1512.04437},
 primaryClass = {astro-ph.EP},
       adsurl = {https://ui.adsabs.harvard.edu/abs/2016MNRAS.457.2877G}
}

@ARTICLE{gilbert,
       author = {{Gilbert}, Gregory J. and {Petigura}, Erik A. and {Entrican}, Paige M.},
        title = "{Planets larger than Neptune have elevated eccentricities}",
      journal = {PNAS},
         year = 2025,
        month = mar,
       volume = {122},
       number = {11},
          eid = {e2405295122},
        pages = {e2405295122},
          doi = {10.1073/pnas.2405295122},
archivePrefix = {arXiv},
       eprint = {2507.07840},
 primaryClass = {astro-ph.EP},
       adsurl = {https://ui.adsabs.harvard.edu/abs/2025PNAS..12205295G}
}

@ARTICLE{goyal,
       author = {{Goyal}, Armaan V. and {Wang}, Songhu},
        title = "{Statistical Reevaluation of the Ultra-short-period Planet Classification Boundary: Smaller Planets within 1 Day, Larger Period Ratios below 2 Days}",
      journal = {\aj},
         year = 2025,
        month = apr,
       volume = {169},
       number = {4},
          eid = {191},
        pages = {191},
          doi = {10.3847/1538-3881/adb487},
archivePrefix = {arXiv},
       eprint = {2502.07773},
 primaryClass = {astro-ph.EP},
       adsurl = {https://ui.adsabs.harvard.edu/abs/2025AJ....169..191G}
}

@ARTICLE{he,
       author = {{He}, Matthias Y. and {Ford}, Eric B. and {Ragozzine}, Darin and {Carrera}, Daniel},
        title = "{Architectures of Exoplanetary Systems. III. Eccentricity and Mutual Inclination Distributions of AMD-stable Planetary Systems}",
      journal = {\aj},
         year = 2020,
        month = dec,
       volume = {160},
       number = {6},
          eid = {276},
        pages = {276},
          doi = {10.3847/1538-3881/abba18},
archivePrefix = {arXiv},
       eprint = {2007.14473},
 primaryClass = {astro-ph.EP},
       adsurl = {https://ui.adsabs.harvard.edu/abs/2020AJ....160..276H}
}

@ARTICLE{izidoro,
       author = {{Izidoro}, Andr{\'e} and {Bitsch}, Bertram and {Raymond}, Sean N. and {Johansen}, Anders and {Morbidelli}, Alessandro and {Lambrechts}, Michiel and {Jacobson}, Seth A.},
        title = "{Formation of planetary systems by pebble accretion and migration. Hot super-Earth systems from breaking compact resonant chains}",
      journal = {\aap},
         year = 2021,
        month = jun,
       volume = {650},
          eid = {A152},
        pages = {A152},
          doi = {10.1051/0004-6361/201935336},
archivePrefix = {arXiv},
       eprint = {1902.08772},
 primaryClass = {astro-ph.EP},
       adsurl = {https://ui.adsabs.harvard.edu/abs/2021A&A...650A.152I}
}

@ARTICLE{lai,
       author = {{Lai}, Dong and {Pu}, Bonan},
        title = "{Hiding Planets behind a Big Friend: Mutual Inclinations of Multi-planet Systems with External Companions}",
      journal = {\aj},
         year = 2017,
        month = jan,
       volume = {153},
       number = {1},
          eid = {42},
        pages = {42},
          doi = {10.3847/1538-3881/153/1/42},
archivePrefix = {arXiv},
       eprint = {1606.08855},
 primaryClass = {astro-ph.EP},
       adsurl = {https://ui.adsabs.harvard.edu/abs/2017AJ....153...42L}
}

@ARTICLE{leleu,
       author = {{Leleu}, Adrien and {Delisle}, Jean-Baptiste and {Burn}, Remo and {Izidoro}, Andr{\'e} and {Udry}, St{\'e}phane and {Dumusque}, Xavier and {Lovis}, Christophe and {Millholland}, Sarah and {Parc}, L{\'e}na and {Bouchy}, Fran{\c{c}}ois and {Bourrier}, Vincent and {Alibert}, Yann and {Faria}, Jo{\~a}o and {Mordasini}, Christoph and {S{\'e}gransan}, Damien},
        title = "{Resonant sub-Neptunes are puffier}",
      journal = {\aap},
         year = 2024,
        month = jul,
       volume = {687},
          eid = {L1},
        pages = {L1},
          doi = {10.1051/0004-6361/202450587},
archivePrefix = {arXiv},
       eprint = {2406.18991},
 primaryClass = {astro-ph.EP},
       adsurl = {https://ui.adsabs.harvard.edu/abs/2024A&A...687L...1L}
}

@ARTICLE{liberles,
       author = {{Liberles}, Benjamin T. and {Dittmann}, Jason A. and {Elardo}, Stephen M. and {Ballard}, Sarah},
        title = "{Variations in the Radius Distribution of Single- and Compact Multiple-transiting Planets}",
      journal = {\aj},
         year = 2024,
        month = aug,
       volume = {168},
       number = {2},
          eid = {92},
        pages = {92},
          doi = {10.3847/1538-3881/ad58da},
archivePrefix = {arXiv},
       eprint = {2312.05809},
 primaryClass = {astro-ph.EP},
       adsurl = {https://ui.adsabs.harvard.edu/abs/2024AJ....168...92L}
}

@ARTICLE{lillo-box,
       author = {{Lillo-Box}, J. and {Gandolfi}, D. and {Armstrong}, D.~J. and {Collins}, K.~A. and {Nielsen}, L.~D. and {Luque}, R. and {Korth}, J. and {Sousa}, S.~G. and {Quinn}, S.~N. and {Acu{\~n}a}, L. and {Howell}, S.~B. and {Morello}, G. and {Hellier}, C. and {Giacalone}, S. and {Hoyer}, S. and {Stassun}, K. and {Palle}, E. and {Aguichine}, A. and {Mousis}, O. and {Adibekyan}, V. and {Azevedo Silva}, T. and {Barrado}, D. and {Deleuil}, M. and {Eastman}, J.~D. and {Fukui}, A. and {Hawthorn}, F. and {Irwin}, J.~M. and {Jenkins}, J.~M. and {Latham}, D.~W. and {Muresan}, A. and {Narita}, N. and {Persson}, C.~M. and {Santerne}, A. and {Santos}, N.~C. and {Savel}, A.~B. and {Osborn}, H.~P. and {Teske}, J. and {Wheatley}, P.~J. and {Winn}, J.~N. and {Barros}, S.~C.~C. and {Butler}, R.~P. and {Caldwell}, D.~A. and {Charbonneau}, D. and {Cloutier}, R. and {Crane}, J.~D. and {Demangeon}, O.~D.~S. and {D{\'\i}az}, R.~F. and {Dumusque}, X. and {Esposito}, M. and {Falk}, B. and {Gill}, H. and {Hojjatpanah}, S. and {Kreidberg}, L. and {Mireles}, I. and {Osborn}, A. and {Ricker}, G.~R. and {Rodriguez}, J.~E. and {Schwarz}, R.~P. and {Seager}, S. and {Serrano Bell}, J. and {Shectman}, S.~A. and {Shporer}, A. and {Vezie}, M. and {Wang}, S.~X. and {Zhou}, G.},
        title = "{TOI-969: a late-K dwarf with a hot mini-Neptune in the desert and an eccentric cold Jupiter}",
      journal = {\aap},
         year = 2023,
        month = jan,
       volume = {669},
          eid = {A109},
        pages = {A109},
          doi = {10.1051/0004-6361/202243879},
archivePrefix = {arXiv},
       eprint = {2210.08996},
 primaryClass = {astro-ph.EP},
       adsurl = {https://ui.adsabs.harvard.edu/abs/2023A&A...669A.109L}
}

@ARTICLE{lissauer11,
       author = {{Lissauer}, Jack J. and {Ragozzine}, Darin and {Fabrycky}, Daniel C. and {Steffen}, Jason H. and {Ford}, Eric B. and {Jenkins}, Jon M. and {Shporer}, Avi and {Holman}, Matthew J. and {Rowe}, Jason F. and {Quintana}, Elisa V. and {Batalha}, Natalie M. and {Borucki}, William J. and {Bryson}, Stephen T. and {Caldwell}, Douglas A. and {Carter}, Joshua A. and {Ciardi}, David and {Dunham}, Edward W. and {Fortney}, Jonathan J. and {Gautier}, III, Thomas N. and {Howell}, Steve B. and {Koch}, David G. and {Latham}, David W. and {Marcy}, Geoffrey W. and {Morehead}, Robert C. and {Sasselov}, Dimitar},
        title = "{Architecture and Dynamics of Kepler's Candidate Multiple Transiting Planet Systems}",
      journal = {\apjs},
         year = 2011,
        month = nov,
       volume = {197},
       number = {1},
          eid = {8},
        pages = {8},
          doi = {10.1088/0067-0049/197/1/8},
archivePrefix = {arXiv},
       eprint = {1102.0543},
 primaryClass = {astro-ph.EP},
       adsurl = {https://ui.adsabs.harvard.edu/abs/2011ApJS..197....8L}
}

@ARTICLE{lissauer14,
       author = {{Lissauer}, Jack J. and {Marcy}, Geoffrey W. and {Bryson}, Stephen T. and {Rowe}, Jason F. and {Jontof-Hutter}, Daniel and {Agol}, Eric and {Borucki}, William J. and {Carter}, Joshua A. and {Ford}, Eric B. and {Gilliland}, Ronald L. and {Kolbl}, Rea and {Star}, Kimberly M. and {Steffen}, Jason H. and {Torres}, Guillermo},
        title = "{Validation of Kepler's Multiple Planet Candidates. II. Refined Statistical Framework and Descriptions of Systems of Special Interest}",
      journal = {\apj},
         year = 2014,
        month = mar,
       volume = {784},
       number = {1},
          eid = {44},
        pages = {44},
          doi = {10.1088/0004-637X/784/1/44},
archivePrefix = {arXiv},
       eprint = {1402.6352},
 primaryClass = {astro-ph.EP},
       adsurl = {https://ui.adsabs.harvard.edu/abs/2014ApJ...784...44L}
}

@ARTICLE{lissauer24,
       author = {{Lissauer}, Jack J. and {Rowe}, Jason F. and {Jontof-Hutter}, Daniel and {Fabrycky}, Daniel C. and {Ford}, Eric B. and {Ragozzine}, Darin and {Steffen}, Jason H. and {Nizam}, Kadri M.},
        title = "{Updated Catalog of Kepler Planet Candidates: Focus on Accuracy and Orbital Periods}",
      journal = {The Planetary Science Journal},
         year = 2024,
        month = jun,
       volume = {5},
       number = {6},
          eid = {152},
        pages = {152},
          doi = {10.3847/PSJ/ad0e6e},
archivePrefix = {arXiv},
       eprint = {2311.00238},
 primaryClass = {astro-ph.EP},
       adsurl = {https://ui.adsabs.harvard.edu/abs/2024PSJ.....5..152L}
}

@ARTICLE{lozovsky,
       author = {{Lozovsky}, M. and {Helled}, R. and {Pascucci}, I. and {Dorn}, C. and {Venturini}, J. and {Feldmann}, R.},
        title = "{Why do more massive stars host larger planets?}",
      journal = {\aap},
         year = 2021,
        month = aug,
       volume = {652},
          eid = {A110},
        pages = {A110},
          doi = {10.1051/0004-6361/202140563},
archivePrefix = {arXiv},
       eprint = {2107.09534},
 primaryClass = {astro-ph.EP},
       adsurl = {https://ui.adsabs.harvard.edu/abs/2021A&A...652A.110L}
}

@ARTICLE{mamajek,
       author = {{Pecaut}, Mark J. and {Mamajek}, Eric E.},
        title = "{Intrinsic Colors, Temperatures, and Bolometric Corrections of Pre-main-sequence Stars}",
      journal = {\apjs},
         year = 2013,
        month = sep,
       volume = {208},
       number = {1},
          eid = {9},
        pages = {9},
          doi = {10.1088/0067-0049/208/1/9},
archivePrefix = {arXiv},
       eprint = {1307.2657},
 primaryClass = {astro-ph.SR},
       adsurl = {https://ui.adsabs.harvard.edu/abs/2013ApJS..208....9P}
}

@ARTICLE{morton,
       author = {{Morton}, Timothy D. and {Bryson}, Stephen T. and {Coughlin}, Jeffrey L. and {Rowe}, Jason F. and {Ravichandran}, Ganesh and {Petigura}, Erik A. and {Haas}, Michael R. and {Batalha}, Natalie M.},
        title = "{False Positive Probabilities for all Kepler Objects of Interest: 1284 Newly Validated Planets and 428 Likely False Positives}",
      journal = {\apj},
         year = 2016,
        month = may,
       volume = {822},
       number = {2},
          eid = {86},
        pages = {86},
          doi = {10.3847/0004-637X/822/2/86},
archivePrefix = {arXiv},
       eprint = {1605.02825},
 primaryClass = {astro-ph.EP},
       adsurl = {https://ui.adsabs.harvard.edu/abs/2016ApJ...822...86M}
}

@INCOLLECTION{mulders24,
       author = {{Mulders}, Gijs D.},
        title = "{Planet Populations as a Function of Stellar Properties}",
    booktitle = {Handbook of Exoplanets},
         year = 2024,
    publisher = {Springer},
          eid = {153},
        pages = {153},
          doi = {10.1007/978-3-319-55333-7_153},
       adsurl = {https://ui.adsabs.harvard.edu/abs/2018haex.bookE.153M}
}

@ARTICLE{muresan,
       author = {{Muresan}, Alexandra and {Persson}, Carina M. and {Fridlund}, Malcolm},
        title = "{Diversities and similarities exhibited by multi-planetary systems and their architectures: I. Orbital spacings}",
      journal = {\aap},
         year = 2024,
        month = dec,
       volume = {692},
          eid = {A122},
        pages = {A122},
          doi = {10.1051/0004-6361/202451353},
archivePrefix = {arXiv},
       eprint = {2410.23399},
 primaryClass = {astro-ph.EP},
       adsurl = {https://ui.adsabs.harvard.edu/abs/2024A&A...692A.122M}
}

@ARTICLE{orell-miquel,
       author = {{Orell-Miquel}, J. and {Nowak}, G. and {Murgas}, F. and {Palle}, E. and {Morello}, G. and {Luque}, R. and {Badenas-Agusti}, M. and {Ribas}, I. and {Lafarga}, M. and {Espinoza}, N. and {Morales}, J.~C. and {Zechmeister}, M. and {Alqasim}, A. and {Cochran}, W.~D. and {Gandolfi}, D. and {Goffo}, E. and {Kab{\'a}th}, P. and {Korth}, J. and {Lam}, K.~W.~F. and {Livingston}, J. and {Muresan}, A. and {Persson}, C.~M. and {Van Eylen}, V.},
        title = "{HD 191939 revisited: New and refined planet mass determinations, and a new planet in the habitable zone}",
      journal = {\aap},
         year = 2023,
        month = jan,
       volume = {669},
          eid = {A40},
        pages = {A40},
          doi = {10.1051/0004-6361/202244120},
archivePrefix = {arXiv},
       eprint = {2211.00667},
 primaryClass = {astro-ph.EP},
       adsurl = {https://ui.adsabs.harvard.edu/abs/2023A&A...669A..40O}
}

@ARTICLE{persson,
       author = {{Persson}, Carina M. and {Knudstrup}, Emil and {Carleo}, Ilaria and {Acu{\~n}a-Aguirre}, Lorena and {Nowak}, Grzegorz and {Muresan}, Alexandra and {Jankowski}, Dawid and {Go{\'z}dziewski}, Krzysztof and {Garc{\'\i}a}, Rafael A. and {Mathur}, Savita and {Palakkatharappil}, Dinil B. and {Borg}, Lina and {Mustill}, Alexander J. and {Barrena}, Rafael and {Fridlund}, Malcolm and {Gandolfi}, Davide and {Hatzes}, Artie P. and {Korth}, Judith and {Luque}, Rafael and {Mart{\'\i}n}, Eduardo L. and {Masseron}, Thomas and {Morello}, Giuseppe and {Murgas}, Felipe and {Orell-Miquel}, Jaume and {Palle}, Enric and {Albrecht}, Simon H. and {Bieryla}, Allyson and {Cochran}, William D. and {Crossfield}, Ian J.~M. and {Deeg}, Hans J. and {Furlan}, Elise and {Guenther}, Eike W. and {Howell}, Steve B. and {Isaacson}, Howard and {Lam}, Kristine W.~F. and {Livingston}, John and {Matson}, Rachel A. and {Matthews}, Elisabeth C. and {Redfield}, Seth and {Schlieder}, Joshua E. and {Seager}, Sara and {Smith}, Alexis M.~S. and {Stassun}, Keivan G. and {Twicken}, Joseph D. and {Van Eylen}, Vincent and {Watkins}, Cristilyn N. and {Weiss}, Lauren M.},
        title = "{TOI-1438: A rare system with two short-period sub-Neptunes and a tentative long-period Jupiter-like planet orbiting a K0V star}",
      journal = {\aap},
         year = 2025,
        month = oct,
       volume = {702},
          eid = {A69},
        pages = {A69},
          doi = {10.1051/0004-6361/202555318},
archivePrefix = {arXiv},
       eprint = {2508.21533},
 primaryClass = {astro-ph.EP},
       adsurl = {https://ui.adsabs.harvard.edu/abs/2025A&A...702A..69P}
}

@ARTICLE{schulze,
       author = {{Schulze}, J.~G. and {Wang}, J. and {Johnson}, J.~A. and {Gaudi}, B.~S. and {Unterborn}, C.~T. and {Panero}, W.~R.},
        title = "{On the Probability That a Rocky Planet's Composition Reflects Its Host Star}",
      journal = {PSJ},
         year = 2021,
        month = jun,
       volume = {2},
       number = {3},
          eid = {113},
        pages = {113},
          doi = {10.3847/PSJ/abcaa8},
archivePrefix = {arXiv},
       eprint = {2011.08893},
 primaryClass = {astro-ph.EP},
       adsurl = {https://ui.adsabs.harvard.edu/abs/2021PSJ.....2..113S}
}

@ARTICLE{spalding,
       author = {{Spalding}, Christopher and {Batygin}, Konstantin},
        title = "{Spin-Orbit Misalignment as a Driver of the Kepler Dichotomy}",
      journal = {\apj},
         year = 2016,
        month = oct,
       volume = {830},
       number = {1},
          eid = {5},
        pages = {5},
          doi = {10.3847/0004-637X/830/1/5},
archivePrefix = {arXiv},
       eprint = {1607.03999},
 primaryClass = {astro-ph.EP},
       adsurl = {https://ui.adsabs.harvard.edu/abs/2016ApJ...830....5S}
}

@ARTICLE{van_eylen18,
       author = {{Van Eylen}, V. and {Agentoft}, Camilla and {Lundkvist}, M.~S. and {Kjeldsen}, H. and {Owen}, J.~E. and {Fulton}, B.~J. and {Petigura}, E. and {Snellen}, I.},
        title = "{An asteroseismic view of the radius valley: stripped cores, not born rocky}",
      journal = {\mnras},
         year = 2018,
        month = oct,
       volume = {479},
       number = {4},
        pages = {4786-4795},
          doi = {10.1093/mnras/sty1783},
archivePrefix = {arXiv},
       eprint = {1710.05398},
 primaryClass = {astro-ph.EP},
       adsurl = {https://ui.adsabs.harvard.edu/abs/2018MNRAS.479.4786V}
}

@ARTICLE{weiss18b,
       author = {{Weiss}, Lauren M. and {Isaacson}, Howard T. and {Marcy}, Geoffrey W. and {Howard}, Andrew W. and {Petigura}, Erik A. and {Fulton}, Benjamin J. and {Winn}, Joshua N. and {Hirsch}, Lea and {Sinukoff}, Evan and {Rowe}, Jason F. and {California Kepler Survey}, The},
        title = "{The California-Kepler Survey. VI. Kepler Multis and Singles Have Similar Planet and Stellar Properties Indicating a Common Origin}",
      journal = {\aj},
         year = 2018,
        month = dec,
       volume = {156},
       number = {6},
          eid = {254},
        pages = {254},
          doi = {10.3847/1538-3881/aae70a},
archivePrefix = {arXiv},
       eprint = {1808.03010},
 primaryClass = {astro-ph.EP},
       adsurl = {https://ui.adsabs.harvard.edu/abs/2018AJ....156..254W}
}

@ARTICLE{xie,
       author = {{Xie}, Ji-Wei and {Dong}, Subo and {Zhu}, Zhaohuan and {Huber}, Daniel and {Zheng}, Zheng and {De Cat}, Peter and {Fu}, Jianning and {Liu}, Hui-Gen and {Luo}, Ali and {Wu}, Yue and {Zhang}, Haotong and {Zhang}, Hui and {Zhou}, Ji-Lin and {Cao}, Zihuang and {Hou}, Yonghui and {Wang}, Yuefei and {Zhang}, Yong},
        title = "{Exoplanet orbital eccentricities derived from LAMOST-Kepler analysis}",
      journal = {PNAS},
         year = 2016,
        month = oct,
       volume = {113},
       number = {41},
        pages = {11431-11435},
          doi = {10.1073/pnas.1604692113},
archivePrefix = {arXiv},
       eprint = {1609.08633},
 primaryClass = {astro-ph.EP},
       adsurl = {https://ui.adsabs.harvard.edu/abs/2016PNAS..11311431X}
}

@ARTICLE{zhu,
       author = {{Zhu}, Wei and {Petrovich}, Cristobal and {Wu}, Yanqin and {Dong}, Subo and {Xie}, Jiwei},
        title = "{About 30\% of Sun-like Stars Have Kepler-like Planetary Systems: A Study of Their Intrinsic Architecture}",
      journal = {\apj},
         year = 2018,
        month = jun,
       volume = {860},
       number = {2},
          eid = {101},
        pages = {101},
          doi = {10.3847/1538-4357/aac6d5},
archivePrefix = {arXiv},
       eprint = {1802.09526},
 primaryClass = {astro-ph.EP},
       adsurl = {https://ui.adsabs.harvard.edu/abs/2018ApJ...860..101Z}
}

@ARTICLE{zink,
       author = {{Zink}, Jon K. and {Christiansen}, Jessie L. and {Hansen}, Bradley M.~S.},
        title = "{Accounting for incompleteness due to transit multiplicity in Kepler planet occurrence rates}",
      journal = {\mnras},
         year = 2019,
        month = mar,
       volume = {483},
       number = {4},
        pages = {4479-4494},
          doi = {10.1093/mnras/sty3463},
archivePrefix = {arXiv},
       eprint = {1901.00196},
 primaryClass = {astro-ph.EP},
       adsurl = {https://ui.adsabs.harvard.edu/abs/2019MNRAS.483.4479Z}
}
\label{ref:lissauer24}
\label{ref:liberles}

\begin{appendix}
\section{Singles and multis}

\begin{table*}
\begin{center}
\caption{Properties of the singles and multis in the three KOI subsets.}
\label{tab:pval_kepler}
\begin{tabular}{clccccccccc}
\hline\hline
\noalign{\smallskip}
FGK & Sub-sample & $N_s$ & $N_m$ & $\widetilde{R_s}$ & $\widetilde{R_m}$ & K-S & A-D & M-W\\
host stars & & & & [$R_\oplus$] & [$R_\oplus$] & p-value & p-value & p-value\\
\noalign{\smallskip}
\hline
\noalign{\smallskip}
\multirow{5}{*}{\hyperref[ref:lissauer24]{L24} subset I} & Full~subset & 1080 & 1055 & 2.19±0.01 & 2.12±0.01 & 0.16±0.11 & 0.04±0.02 & 0.25±0.09\\ & Excl.~HJs\tablefootmark{a} & 1029 & 1050 & 2.14±0.01 & 2.12±0.01 & 0.43±0.18 & $\geq0.25$ & 0.62±0.15\\
& \mbox{$R < 4\, R_\oplus$} & 959 & 961 & 2.05±0.02 & 2.03±0.01 & 0.40±0.17 & $\geq0.25$ & 0.87±0.10\\
& \mbox{$R \geq 4\, R_\oplus$}\tablefootmark{b} & 121 & 94 & 8.32±0.24 & 6.41±0.22 & 0.001±0.002 & $\leq10^{-3}$ & 0.001±0.002\\
\hline
\noalign{\smallskip}
\multirow{5}{*}{\hyperref[ref:lissauer24]{L24} subset II} & Full~subset & 1080 & 1055 & 2.19±0.01 & 2.13±0.01 & 0.24±0.10 &  0.04±0.02 &  0.36±0.12\\ & Excl.~HJs\tablefootmark{a} & 1031 & 1053 & 2.13±0.01 & 2.13±0.01 & 0.26±0.14 & $\geq0.25$ & 0.45±0.13\\
& \mbox{$R < 4\, R_\oplus$} & 966 & 969 & 2.05±0.01 & 2.04±0.01 & 0.27±0.16 & $\geq0.25$ & 0.81±0.14\\
& \mbox{$R \geq 4\, R_\oplus$}\tablefootmark{b} & 114 & 86 & 8.66±0.22 & 6.31±0.14 & $<10^{-3}$ & $\leq10^{-3}$ & $<10^{-4}$\\
\hline
\noalign{\smallskip}
\multirow{5}{*}{subset III} & Full~subset & 1080 & 1055 & 2.11±0.02 & 2.16±0.01 & 0.27±0.15 & 0.18±0.05 & 0.43±0.14\\ 
This work's & Excl.~HJs\tablefootmark{a} & 1036 & 1053 & 2.05±0.02 & 2.16±0.01 & 0.07±0.06 & 0.02±0.01 & 0.02±0.01\\
& \mbox{$R < 4\, R_\oplus$} & 979 & 964 & 1.99±0.02 & 2.05±0.01 & 0.19±0.12 & 0.18±0.06 & 0.18±0.09\\
& \mbox{$R \geq 4\, R_\oplus$}\tablefootmark{b} & 101 & 91 & 7.70±0.23 & 5.98±0.14 & 0.01±0.01 & 0.001±0.002 & 0.001±0.002\\
\hline
\end{tabular}
\end{center}
\tablefoot{Comparisons of the confirmed singles and multis orbiting FGK stars in each of the three KOI subsets (Sect.~\ref{sect:subsets}). $s$ and $m$ stand for singles and multis, respectively, while $N$ and $\widetilde{R}$ represent the number and the median radii of the planets in each sample, respectively. The p-values indicate the significance of the differences between $R_s$ and $R_m$ in each sample, as measured with three tests: the Kolmogorov–Smirnov (K-S), the two-sample Anderson-Darling (A-D), and the Mann–Whitney (M-W) U test. All the reported $\pm1\sigma$-uncertainties are computed via Monte Carlo simulations (Sect.~\ref{sect:radii}).}
\tablefoottext{a}{This sub-sample comprises all the singles and multis in the sample, after removing the hot Jupiters.}
\tablefoottext{b}{This sub-sample contains the giants with $R \geq 4\, R_\oplus$, excluding the hot Jupiters.}
\end{table*}

\subsection{Ultra-short-period planets}

Within the small-planet population, ultra-short-period (USP) planets constitute a special planet category, as listed in Tables~\ref{tab:pct_pltypes} and \ref{tab:pct_singles}, and have been found to reside primarily in single-planet systems \citep{weiss18b, goyal}. 
In the FGK sample, singles account for $61 \pm5\%$ of all the USP planets with $P < 1$~day and $R < 4\, R_\oplus$. 
For the USP sub-sample with FGK~host stars, all three comparison tests, as outlined in Sect.~\ref{sect:comparison}, indicate that both the radius and the orbital period distributions of the singles are statistically indistinguishable from those of the multis. 
Notably, there are 83 FGK systems in which the innermost planet is a USP planet with \mbox{$P < 1$ day} and \mbox{$R < 2\, R_\oplus$}. Singles account for 49 (\mbox{$59 \pm5\%$}) of these systems and, thus, constitute a lower fraction of the observed USP systems than previously reported by \citet{goyal}, who also included systems with USP candidate planets. 
These numbers indicate that while more than 50\% of the confirmed USPs are singles, this percentage increases when USP candidates are also included in the sample.

\subsection{Statistical tests for comparisons} \label{apx:b}
 
As reported in Sect.~\ref{sect:radii} and Table~\ref{tab:pval}, we assessed the significance of the differences between the radii of the singles and multis by employing three statistical tests: a Kolmogorov–Smirnov (K-S) test, a two-sample Anderson-Darling (A-D) test, and a Mann–Whitney (M-W) U test. They are non-parametric tests and do not require or assume the data to follow a specific probability distribution. The null hypothesis posits that the two samples, which are being compared, can be drawn from the same population.
The K-S test mainly compares the overall shapes of the two distributions, while the M-W test focuses on the differences in their averages, and the A-D test is most sensitive to differences in the tails of the two distributions. Compared to the other two tests, the A-D implementation in SciPy employed in this work outputs a p-value in the interval from 0.01 to 0.25.\footnote{SciPy k-sample test: \url{https://docs.scipy.org/doc/scipy/reference/generated/scipy.stats.anderson_ksamp.html}}
The main results from our comparisons between the multis and singles in several sub-samples are presented in Table~\ref{tab:pval}.
We note that the K-S test p-values given in Figs.~\ref{fig:mesh_fgk} and \ref{fig:mesh_m} differ slightly from those reported in Table~\ref{tab:pval} since they have not been subjected to Monte Carlo simulations, as explained in Sect.~\ref{sect:radii}.

\end{appendix}

\end{document}